\documentclass[prx,amsmath,amssymb,aps,floatfix,twocolumn]{revtex4-2}

\usepackage{graphicx}
\usepackage{bm}
\usepackage[colorlinks=true]{hyperref}
\usepackage[capitalise]{cleveref}
\usepackage{braket}
\usepackage{color}
\usepackage[export]{adjustbox}

\newcommand{\tr}{\text{tr}}
\newcommand{\re}{{\rm e}}
\newcommand{\ri}{{\rm i}}

\begin{document}

%%%%%%%%%%%%%%%%%%%%%%%%%%%%%%%%%%%%%%%%%%%%%%%%%%%%%%%%%%%%
\title{Quantum Skyrmion Lattices in Heisenberg Ferromagnets}
%%%%%%%%%%%%%%%%%%%%%%%%%%%%%%%%%%%%%%%%%%%%%%%%%%%%%%%%%%%%
%
\def\aff{Department of Physics and Materials Science, University of Luxembourg, L-1511 Luxembourg, Luxembourg}
\author{Andreas Haller}
\email{andreas.haller@uni.lu}
\affiliation{\aff}
\author{Solofo Groenendijk}
\affiliation{\aff}
\author{Alireza Habibi}
\affiliation{\aff}
\author{Andreas Michels}
\affiliation{\aff}
\author{Thomas L. Schmidt}
\email{thomas.schmidt@uni.lu}
\affiliation{\aff}
\date{\today}
\begin{abstract}
    Skyrmions are topological magnetic textures that can arise in non-centrosymmetric ferromagnetic materials.
    In most systems experimentally investigated to date, skyrmions emerge as classical objects.
    However, the discovery of skyrmions with nanometer length scales has sparked interest in their quantum properties.
    Here, we simulate the ground states of two-dimensional spin-$1/2$ Heisenberg lattices with Dzyaloshinskii-Moriya interactions and discover a broad region in the zero-temperature phase diagram which hosts quantum skyrmion lattices.
    We argue that the quantum skyrmion lattice phase can be detected experimentally in the magnetization profile via local magnetic polarization measurements as well as in the spin structure factor measurable via neutron scattering experiments.
    Finally, we explore the resulting quantum skyrmion state, analyze its real-space polarization profile and show that it is a non-classical state featuring entanglement between quasiparticle and environment mainly localized near the boundary spins of the skyrmion.
\end{abstract}
\maketitle
%
%%%%%%%%%%%%%%%%%%%%%%%%%%%%%%%%%%%%%%%
\section{\label{sec:intro}Introduction}
%%%%%%%%%%%%%%%%%%%%%%%%%%%%%%%%%%%%%%%
Magnetic skyrmions are vortex-like quasiparticles characterized by a nontrivial topological invariant in real space~\cite{bogdanov1989,bogdanov1994,rosslerSpontaneousSkyrmionGround2006,neubauer2009}.
These states are typically found in non-centrosymmetric ferromagnets in a certain range of external magnetic field and temperature, and are stabilized by an antisymmetric spin exchange energy, termed Dzyaloshinskii-Moriya interaction (DMI)~\cite{bogdanov1989}.
After their first detection in a magnetic system by a neutron diffraction experiment in 2009~\cite{muhlbauerSkyrmionLatticeChiral2009}, and a full microscopic to\-mo\-graphy by electron microscopy in 2010~\cite{yu2010}, intense follow-up studies revealed intriguing dynamical properties, rendering skyrmions potentially useful for memory and computing devices~\cite{Jonietz2010,Fert2013,wiesendangerNanoscaleMagneticSkyrmions2016,EverschorSitte2018,Mandru2020}.
Usually, the skyrmions encountered in these systems arise from thermal fluctuations and extend over length scales that are much larger than the interatomic distance and thus behave like classical objects.
Other possibilities to create skyrmions are through suitable DC current devices, such as those proposed in Refs.~\cite{Stier2017,EverschorSitte2017}.
However, smaller skyrmions do exist~\cite{Heinze_2011} and have already created interest in possible quantum properties of skyrmions.
Several works have predicted the quantum behavior of skyrmions by using classical magnetic textures as a starting point and studying quantum corrections in the semiclassical regime~\cite{ochoaQuantumSkyrmionics2019, psaroudakiQuantumDynamicsSkyrmions2017, takashimaQuantumSkyrmionsTwodimensional2016,RoldanMolina2015,schutteMagnonskyrmionScatteringChiral2014}.

Beyond this semiclassical limit, some works have indicated that quantum analogs of classical skyrmions might exist in spin systems.
The authors of Ref.~\cite{Janson2014} used a multiscale approach to demonstrate that mesoscopic magnetization vortices are stabilized by quantum fluctuations~\cite{Seki2012,RoldanMolina2015}, which suggests the possibility of inherently quantum-mechanical counterparts of these states at zero temperature.
So far, attempts to classify skyrmion excitations with sizes comparable to the interatomic spacing have been made in frustrated quantum lattice systems~\cite{Lohani2019} and ferromagnetic lattices with DMI~\cite{Sotnikov2021,sieglControlledCreationQuantum2021}.
Several geometries have been studied to understand the quantum analogs of classical skyrmions, and quantitative results have been obtained by numerical diagonalization of the Hamiltonian~\cite{Lohani2019,Sotnikov2021,sieglControlledCreationQuantum2021}.
Since the dimension of the quantum Hamiltonian scales exponentially with the number of lattice sites, such exact diagonalization (ED) strategies are limited to small system sizes containing at most $\approx 30$ spin-$1/2$ sites (without exploiting symmetries).
Although DMI interactions are among the most popular to investigate the formation of classical skyrmion phases, their quantum analogs are analytically hard to handle and quantitative results beyond system sizes amenable for ED are still lacking.
As an alternative route to quantum skyrmions that avoids DMI, frustrated spin lattice systems were studied in Ref.~\cite{Lohani2019}.
Using ED for small systems and analytical spin wave theory, the authors identified skyrmions with magnon bound states and developed a phenomenological theory based on a trial wave function.

Here, we use the density matrix renormalization group (DMRG) algorithm to explore ferromagnetic phases of quantum spin-$1/2$ Heisenberg models with DMI and uniaxial anisotropy.
As our main result, we discover a zero-temperature quantum phase with a nontrivial magnetic spin texture that signals an emergent quantum skyrmion lattice.
This phase was previously overlooked because it appears only beyond a critical system size which for realistic parameters is larger than the system sizes amenable to ED.
We identify three ferromagnetic phases that can be directly observed and distinguished in the space-resolved magnetization profile.
Furthermore, we argue that the polarization gives access to the zero-temperature phase diagram of the model under investigation.
Contrary to similar quasiparticles found in frustrated lattices~\cite{Lohani2019} or quantum skyrmions embedded in a classical magnet~\cite{sieglControlledCreationQuantum2021}, we show that the skyrmion lattice phase reported in this letter emerges from entangled spin-$1/2$ pairs, which bear witness to a genuine quantum mechanical origin without classical analog, a feature that may pave the way towards a microscopic description of skyrmion qubits used for realizing quantum logic elements based on nanoscale devices~\cite{Psaroudaki2021}

%%%%%%%%%%%%%%%%%%%%%%%%%%%%%%%%
\section{\label{sec:model}Model}
\begin{figure}[t]
    \includegraphics[width=\columnwidth]{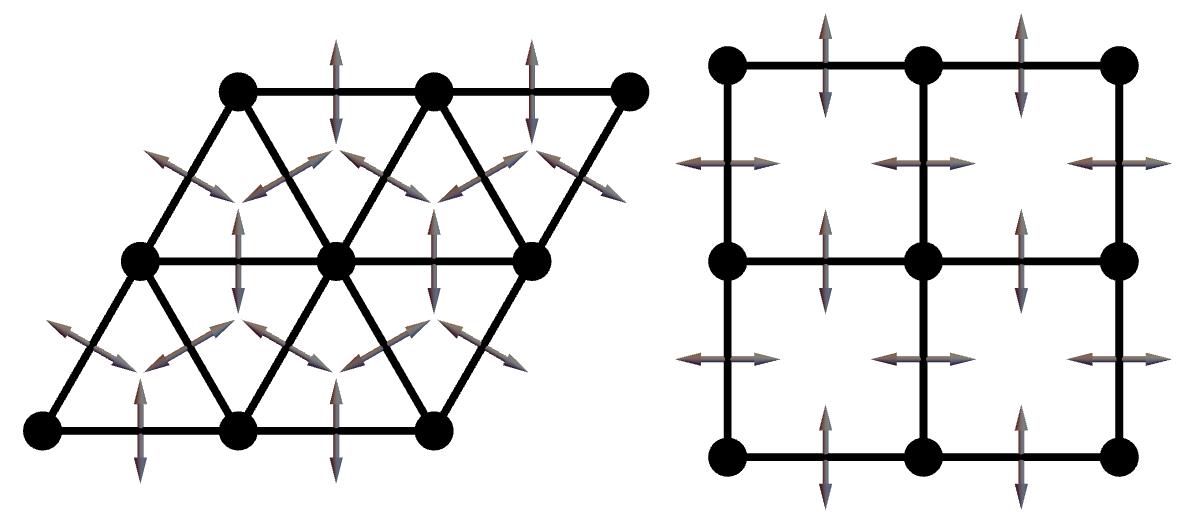}
    \caption{DMI vector pattern $\bm D_{\pm \bm a_i}$ for nearest-neighbor interactions on triangular and square lattices.}
    \label{fig:model}
\end{figure}

%%%%%%%%%%%%%%%%%%%%%%%%%%%%%%%%
We study the zero-temperature phase diagram of a quantum spin-$1/2$ Heisenberg model with DMI and external magnetic field. The Hamiltonian reads:
\begin{align}
    \hat H &= \frac12\sum_{\braket{\bm r, \bm r'}}\left[J\hat{\bm S}_{\bm r} \cdot \hat{\bm S}_{\bm r'} + {\bm D}_{{\bm r}'-{\bm r}}\cdot \left( \hat{\bm S}_{\bm r}\times\hat{\bm S}_{\bm r'}\right)\right]\nonumber\\
    &+ \sum_{\bm r}{\bm B} \cdot \hat{\bm S}_{\bm r} ,\
    \label{eq:H}
\end{align}
where $\hat{\bm S}_{\bm r}=\hbar\hat{\bm\sigma}_{\bm r}/2$, with Pauli matrices $\hat\sigma_{\alpha,\bm r}$ for $\alpha \in \{x,y,z\}$, denotes a spin-$1/2$ operator at position $\bm r$.
$J<0$ is the ferromagnetic exchange coupling strength, ${\bm D}_{\bm r'-\bm r}$ is the DMI vector, and $\bm B = B \hat{\bm e}_z$ denotes the applied magnetic field along the $z$~axis.
The notation $\braket{\bm r,\bm r'}$ implies a sum over all pairs of nearest-neighbor lattice sites.
In \cref{sec:skx}, we will also consider the impact of a uniaxial magnetic anisotropy with strength $K$,
\begin{equation}
    {\hat H}_K = \frac12\sum_{\braket{\bm r, \bm r'}}K{\hat S}^z_{\bm r}{\hat S}^z_{\bm r'}.
    \label{eq:HK}
\end{equation}
We solve the above Hamiltonian numerically by means of matrix product state (MPS) simulations on different two-dimensional Bravais lattices consisting of lattice sites ${\bm r} = \sum_in_i{\bm a_i}$ spanned by basis vectors $\bm a_{1,2}$ with $a_{z,i} = 0$.
For details on the implementation and on the 2D-1D mapping required for using MPS, we refer to the Appendix and Ref.~\cite{Haller2021Zenodo}.
The DMI vectors read
\begin{equation}
    \bm D_{\bm r'-\bm r} = D\hat{\bm e}_z\times({\bm r'-\bm r}),
\end{equation}
with the positive DMI vector amplitude $D>0$.
For triangular and square lattices, we depicted the orientation of the DMI vectors in~\cref{fig:model}.

Note that we apply the external magnetic field $\bm B$ parallel to the lattice plane normal $\hat{\bm e}_z$.
Without loss of generality, we assume negative values $B<0$ such that field-polarized spins are eigenstates of $\hat S^z_{\bm r}$ with eigenvalue $+1/2$ and therefore align with the plane normal.
For convenience, we will express interatomic distances in units of the lattice constants $a_i=|{\bm a}_i|$ and energies in units of $D$, with $\hbar=1$.

The Hamiltonian~\eqref{eq:H} may be seen as the quantum counterpart of typical classical spin models which give rise to magnetic skyrmions~\cite{bogdanov1989,bogdanov1994,rosslerSpontaneousSkyrmionGround2006,neubauer2009}.
In this work, we discuss the emergence of quantum skyrmions and quantum skyrmion lattices in two-dimensional triangular and square lattices at zero temperature and with different boundary shapes (see \cref{fig:three_phases_different_BC}).
\medskip

%%%%%%%%%%%%%%%%%%%%%%%%%%%%%%%%%%%%%%%%%%%%%%%%%%%%
\section{\label{sec:sk}Individual quantum skyrmions}
%%%%%%%%%%%%%%%%%%%%%%%%%%%%%%%%%%%%%%%%%%%%%%%%%%%%
For small system diameters ($L \lesssim 5a$), the exact eigenvalues and eigenvectors of the full Hamiltonian can be computed numerically, for instance by the Lanczos or Arnoldi algorithms, and we first cross-checked our own exact diagonalization (ED) codes by reproducing the results of Ref.~\cite{Sotnikov2021}.
To ensure the correctness of our findings, we further compared expectation values computed from the matrix product states (MPS) obtained by DMRG with ED results for all system sizes amenable for ED and found quantitative agreement up to a self-imposed accuracy $\delta$.
This shows that finite-size ground states of the spin-$1/2$ Hamiltonian can be approximated faithfully with MPS (for more details, see \cref{sec:MPS} and \cite{Haller2021Zenodo}).
In fact, the reliability of DMRG for 2D spin-$1/2$ quantum Heisenberg models has already been demonstrated in numerous works, with a strong bias towards frustrated antiferromagnets and spin liquids, which are prime examples of the most demanding systems to simulate numerically due to the presence of topological order and long-range entanglement~\cite{Capriotti2004,White2007,Kallin2009,Yan2011,Depenbrock2012,Jiang2012natcom,Nishimoto2013,Zhu2013,Gong2014,He2014,Ramos2014,Kolley2015,Shinjo2015,Hu2015,Iqbal2016,Morita2016,Saadatmand2016,He2017,Capponi2017,Saadatmand2017,Gohlke2017,Wang2018,Chen2018,Verresen2018,Haghshenas2018,Gong2019,Hu2019,Dong2020,Schaefer2020,Hagymasi2021,liu2021gapless}.
In this work, we focus on the computationally less demanding scenario of magnetically ordered phases hosting skyrmions and field-polarized (FP) states in the regime $J<0$.

\begin{figure}[h!]
    \includegraphics[width=\columnwidth]{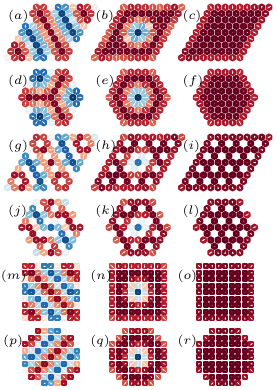}
    \caption{Local polarization $m_{z,\bm r}$ (in color) of the ground states of the three different phases as obtained by DMRG simulations of \cref{eq:H}. Arrows depict the direction and size of the magnetization components $m_{x/y,\bm r}$ perpendicular to the external field. We study different triangular $(a)$-($l$) and square $(m)$-$(r)$ systems with regular boundaries (odd rows) and circular boundaries (even rows). The parameters used are $J=-D/2$ and $K=0$, with a varying external field $B=-0.1D$ (first column), $B=-0.5D$ (second column) and $B=-1.0D$ (third column). We find quantum skyrmions for system sizes larger than a critical diameter $L\approx8a$, irrespective of the lattice symmetries and boundary conditions.}
    \label{fig:three_phases_different_BC}
\end{figure}

Beyond a critical system diameter of $L\approx8a$, we find values of $D$ and $B$ for which the ground state of the Hamiltonian $\hat{H}$ hosts skyrmion-like spin textures confined in the interior (bulk) of the lattice, which we display in \cref{fig:three_phases_different_BC}.
We checked that these skyrmion wave functions correspond to approximate eigenstates of the Hamiltonian by computing the energy variance $\varepsilon$ and performed a linear extrapolation towards results without numerical errors (see \cref{sec:MPS} for a discussion).

\begin{figure}[ht]
    \includegraphics{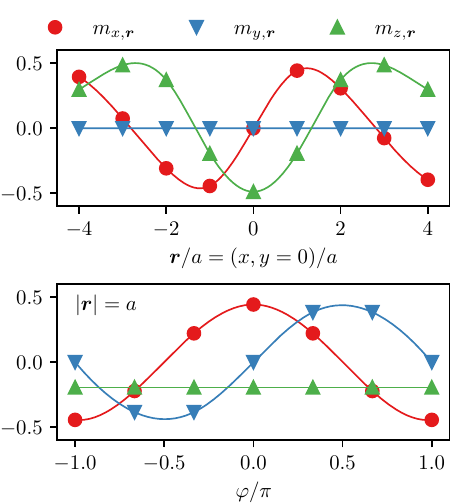}
    \caption{Components of the local polarization $\bm m_{\bm r}$ of the centered skyrmion ground state depicted in \cref{fig:three_phases_different_BC}(e). From the radial winding at fixed $y=0$ (upper panel) the radius of the quasiparticle can be estimated to $r_0\approx3a$. The azimuthal projection (lower panel) at fixed $|\bm r|=a$ reveals a sinusoidal winding of the components $m_{x,y}$, with a phase difference of $\pi$.}
    \label{fig:polarization_winding}
\end{figure}

\begin{figure}[hb]
    \includegraphics[width=\columnwidth]{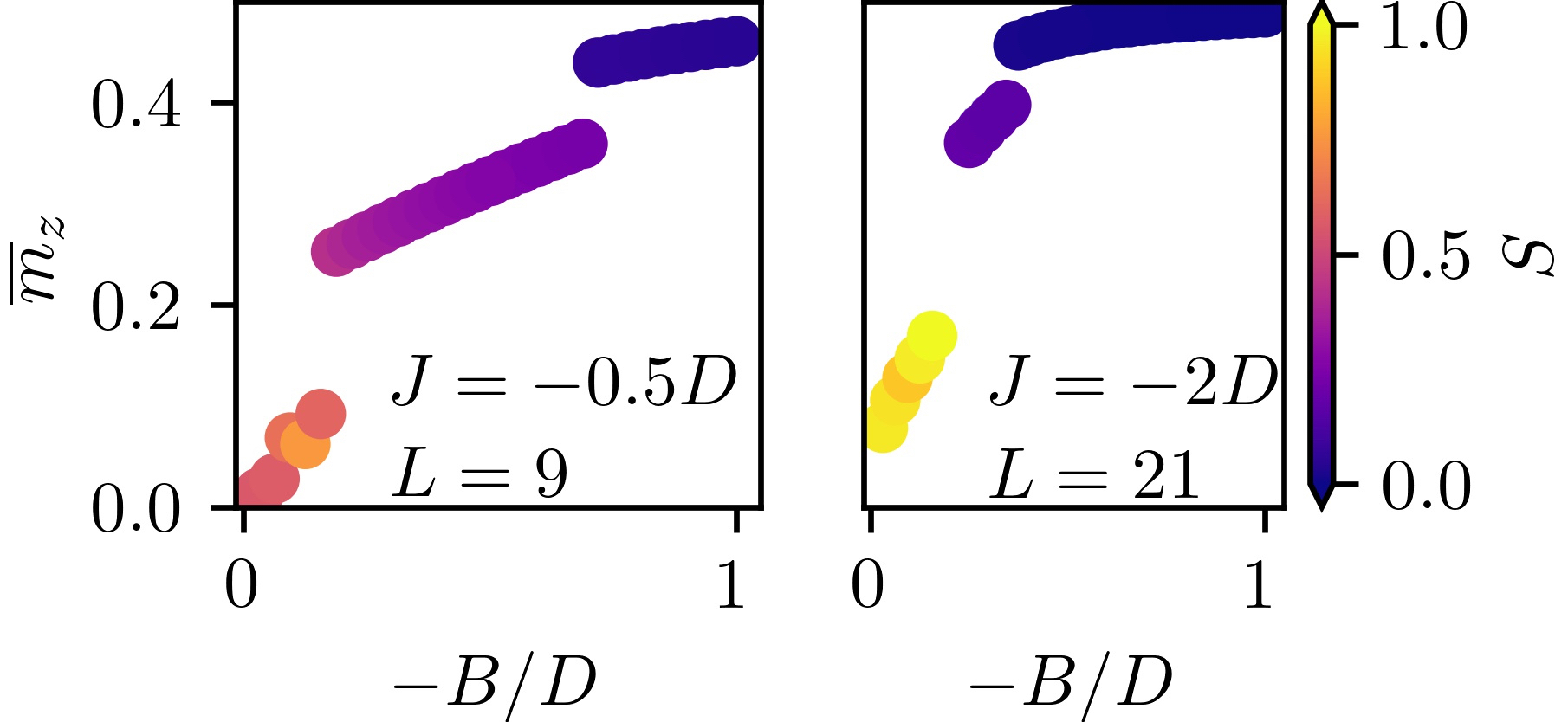}
    \caption{Average polarization $\overline{m}_z$ as a function of the external field strength $B$. The ferromagnetic exchange interaction (without anisotropy, $K=0$) is fixed to $J=-0.5 D$ (system diameter $9$) and $J=-2D$ (system diameter $21$) as indicated in the figure. The color scale indicates the maximum entanglement entropy $S$ (see text).}
    \label{fig:polarization_plateaus}
\end{figure}

We compute the components of the spin magnetization, which are local expectation values ${\bm m}_{\bm r}=\braket{\hat{\bm S}_{\bm r}}$.
Since $m_{z,\bm r}$ is parallel to the external field, we call this magnetization component the polarization. For $B\approx J=-D/2$, the local spin profiles yield magnetization textures similar to those obtained for classical skyrmion configurations of the N\'eel (hedgehog) type: the central spin is polarized opposite to the applied magnetic field, and the spins wind radially from the center towards the periphery.
In \cref{fig:three_phases_different_BC}, we depict the polarization along the field ($m_{z,\bm r}$) using a color scale and the in-plane polarization ($m_{x/y,\bm r}$) by arrows.
More detailed radial and angular distributions of panel \cref{fig:three_phases_different_BC}(e) are depicted in \cref{fig:polarization_winding}.
In \cref{fig:polarization_plateaus}, we show that the average polarization $\overline{m}_z = \frac1N\sum_{\bm r}m_{z,\bm r}$, as a function of the external field, yields three disconnected regions uniquely associated with the three phases of \cref{eq:H}.

We estimate the size of an individual quantum skyrmion as the number of lattice sites over which the polarization changes its orientation once and the components orthogonal to the external field vanish. The radius can then be read out from \cref{fig:three_phases_different_BC,fig:polarization_winding} and results in $r_0\approx 3a$ for $J=-0.5D$ and $K=0$.
These quantum skyrmion (SK) ground states occur not only for the fine-tuned parameters presented in \cref{fig:three_phases_different_BC} but in a wide range of intermediate values of the magnetic field.
Furthermore, the emergence of individual skyrmions for small lattices is largely independent of the lattice geometry.
While the effect of boundaries cannot be neglected for the small systems considered here, we verified that the size of an individual skyrmion is neither affected by the boundary conditions (see \cref{fig:three_phases_different_BC}) nor the system diameter (see \cref{fig:lattices}).

As we show in \cref{fig:three_phases_different_BC}, for small magnetic fields, the system's ground state is a helical spin spiral (HS) state.
This is characterized by a degenerate ground state, in which each possible ground state features an oscillation of the magnetization along a symmetry axis of the lattice.
The large ground state degeneracy makes the spin spiral phase notoriously difficult to simulate for tensor network states.
In contrast, for large magnetic fields, the system reaches a ferromagnetic state, where all spins are polarized parallel to the external magnetic field.
The bulk of the field-polarized ferromagnet is devoid of entanglement and can thus be most efficiently approximated by an MPS.
For parameters that result in SK ground states, DMRG reliably converges within a few dozen sweeps and yields excellent MPS approximations with maximum truncation error $\Delta\rho\approx10^{-6}$, even for small bond dimensions $M=32$~\cite{Haller2021Zenodo}.

We want to stress that the fine-tuned regime $J=-0.5D$ is not necessary to enter the skyrmion lattice phase.
By increasing the ferromagnetic exchange coupling to a value $J=-2D$, we obtain qualitatively similar results, which we present in \cref{fig:polarization_plateaus}.
Note, however, that the corresponding system diameter for $J=-2D$ is dramatically increased compared to $J=-0.5D$.
We attribute this to the fact that the skyrmion radius is controlled by the ratio $J/D$.

% place figures a little early to be displayed at the appropriate page
%
\begin{figure}[ht]
    \includegraphics[width=\columnwidth]{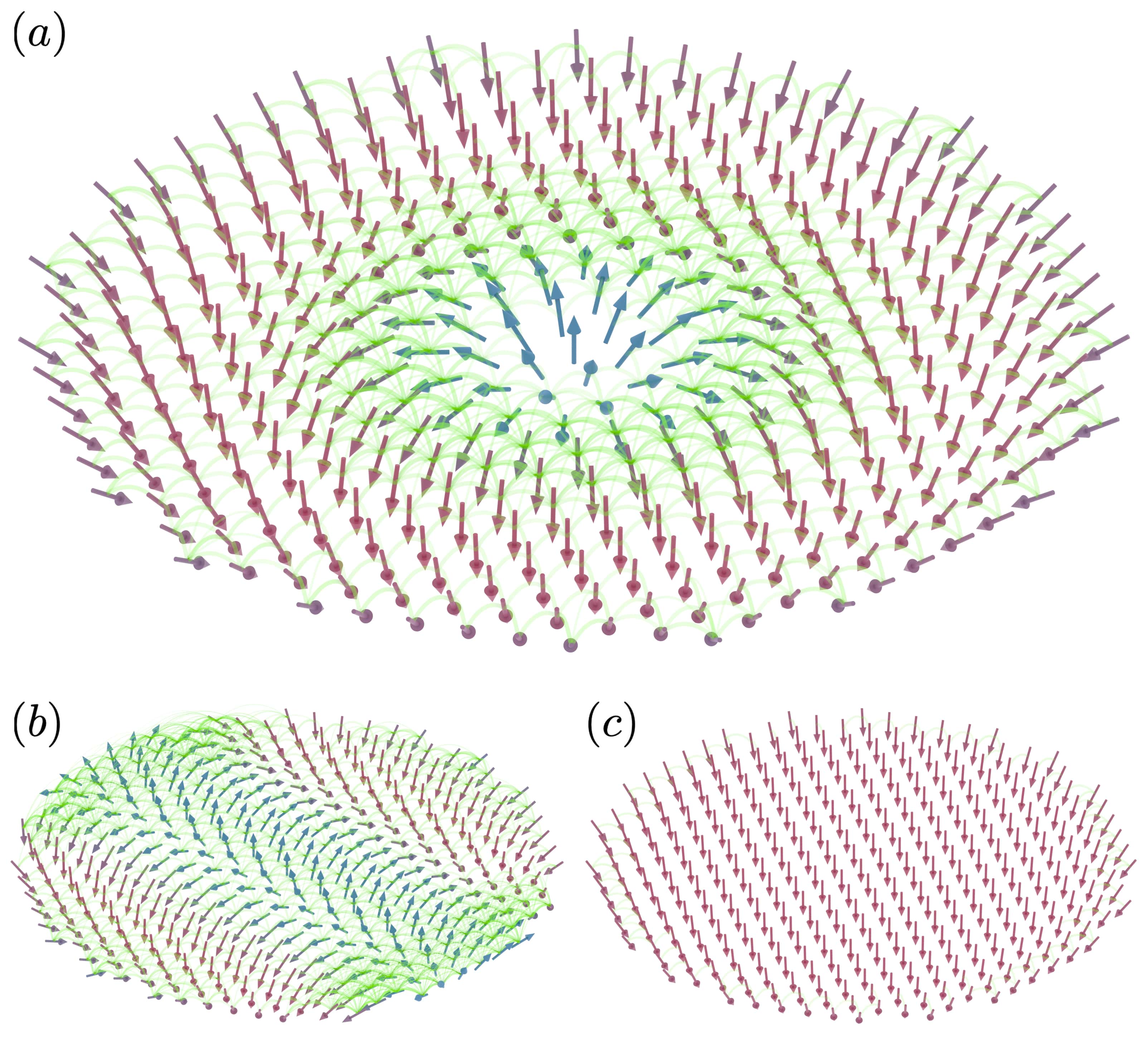}
    \caption{Concurrence sketch of representative ground states in the three phases without anisotropy $K=0$ and $J=-2D$. The system diameter is $L=21$. We plot the local polarization as colored arrows and the concurrence by green lines connecting pairs of sites. In panel $(a)$, we present a SK state at $B=-D/4$, in $(b)$ a HS ground state for $B=-D/16$ and in $(c)$ a FP state at $B=-3D/4$.}
    \label{fig:concurrence}
\end{figure}

%%%%%%%%%%%%%%%%%%%%%%%%%%%%%%%%%%%%%%%%%%%%%%
\section{\label{sec:entanglement}Entanglement}
%%%%%%%%%%%%%%%%%%%%%%%%%%%%%%%%%%%%%%%%%%%%%%

\begin{table}[ht]
    \centering
    \begin{tabular}{| c || c | c | c |}
        \hline
        state & n.n. & n.n.n. & 3rd n.n.\\
        \hline\hline
        HS
        &
        \includegraphics[valign=c,width=0.282\columnwidth]{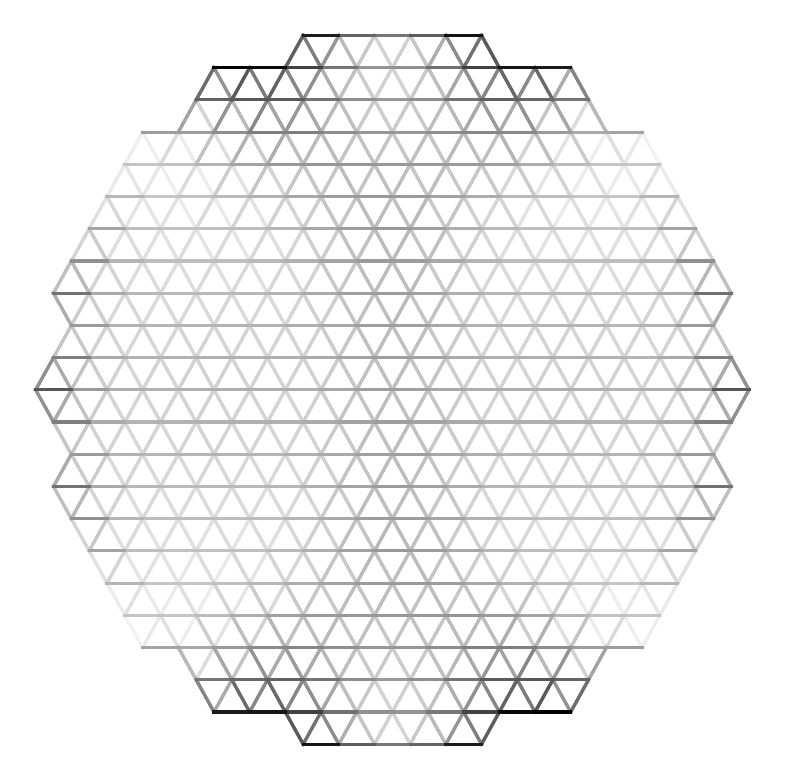}
        &
        \includegraphics[valign=c,width=0.282\columnwidth]{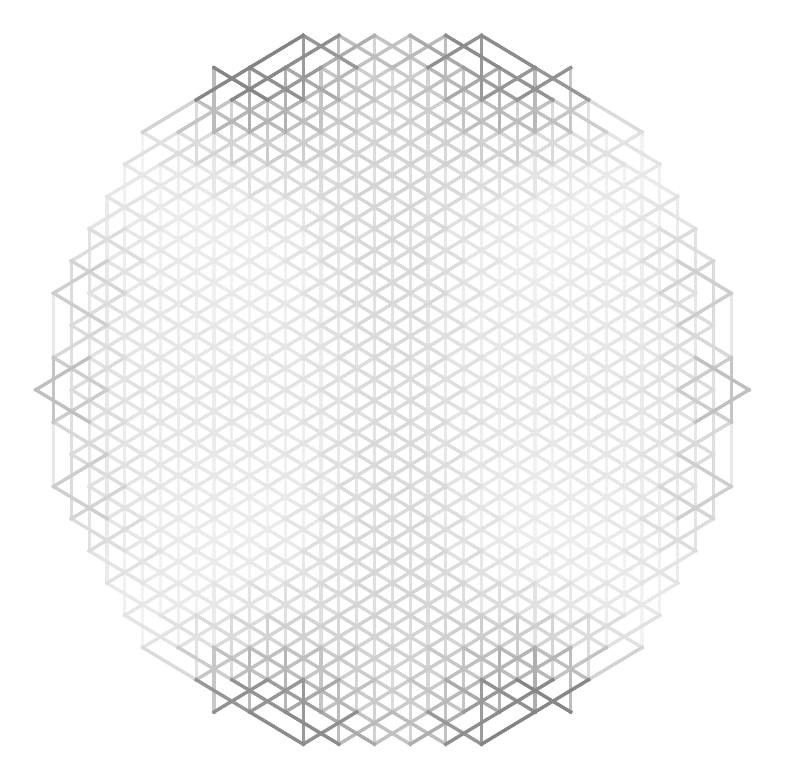}
        &
        \includegraphics[valign=c,width=0.282\columnwidth]{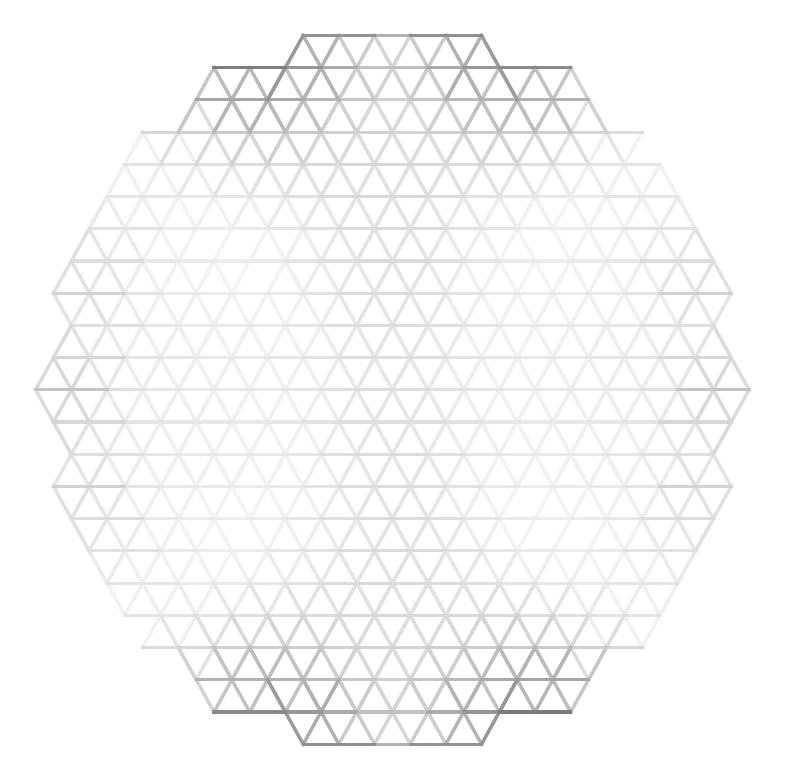}
        \\
        \hline
        SK
        &
        \includegraphics[valign=c,width=0.282\columnwidth]{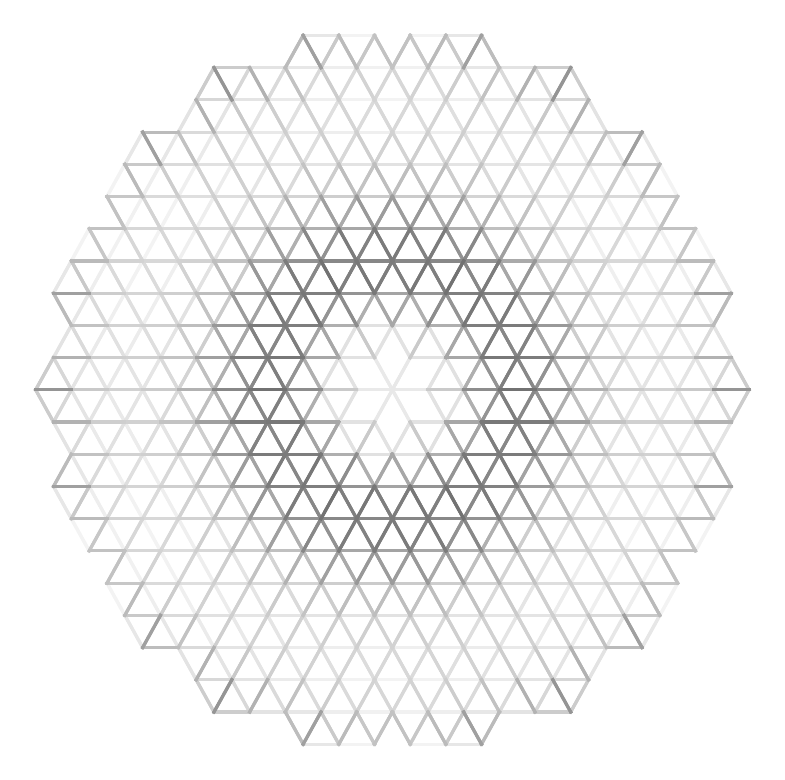}
        &
        \includegraphics[valign=c,width=0.282\columnwidth]{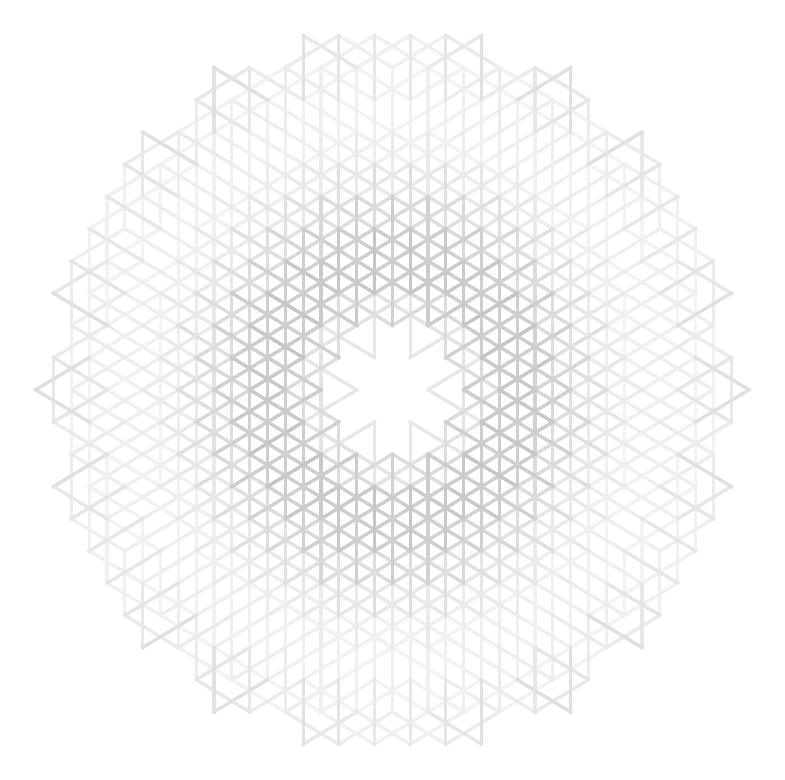}
        &
        \includegraphics[valign=c,width=0.282\columnwidth]{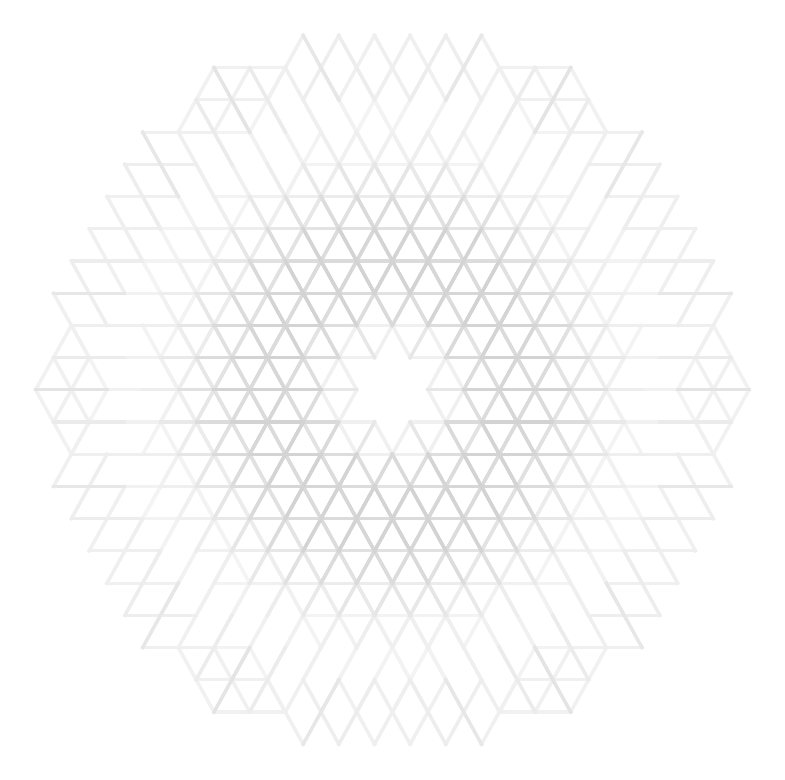}
        \\
        \hline
        FP
        &
        \includegraphics[valign=c,width=0.282\columnwidth]{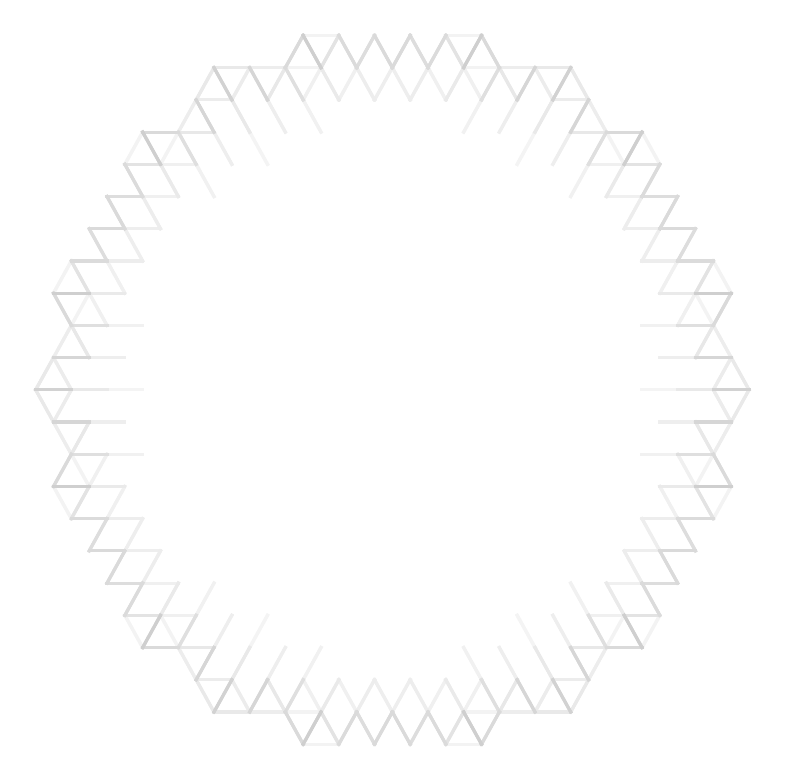}
        &
        \includegraphics[valign=c,width=0.282\columnwidth]{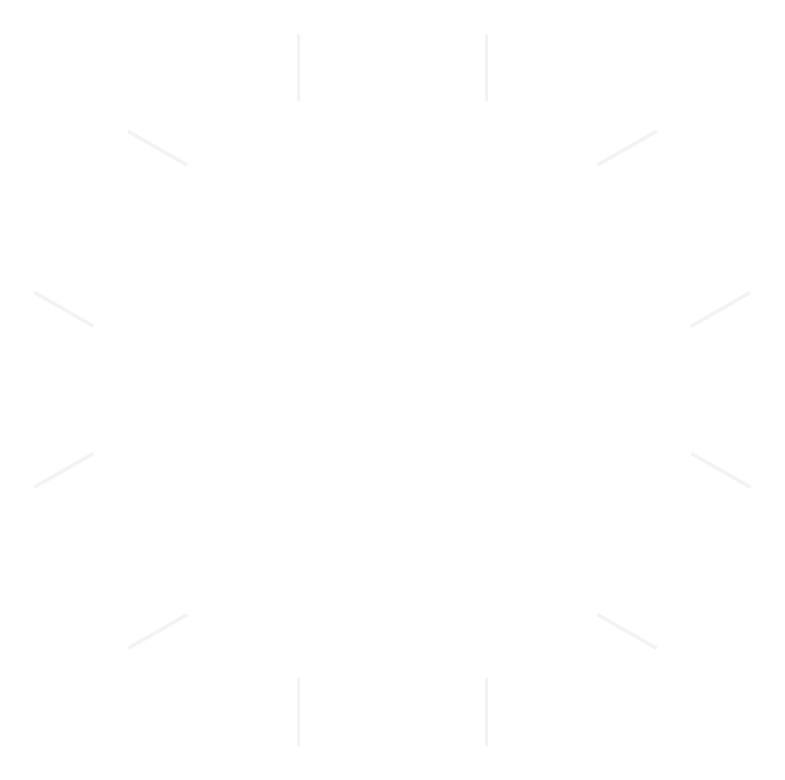}
        &
        \includegraphics[valign=c,width=0.282\columnwidth]{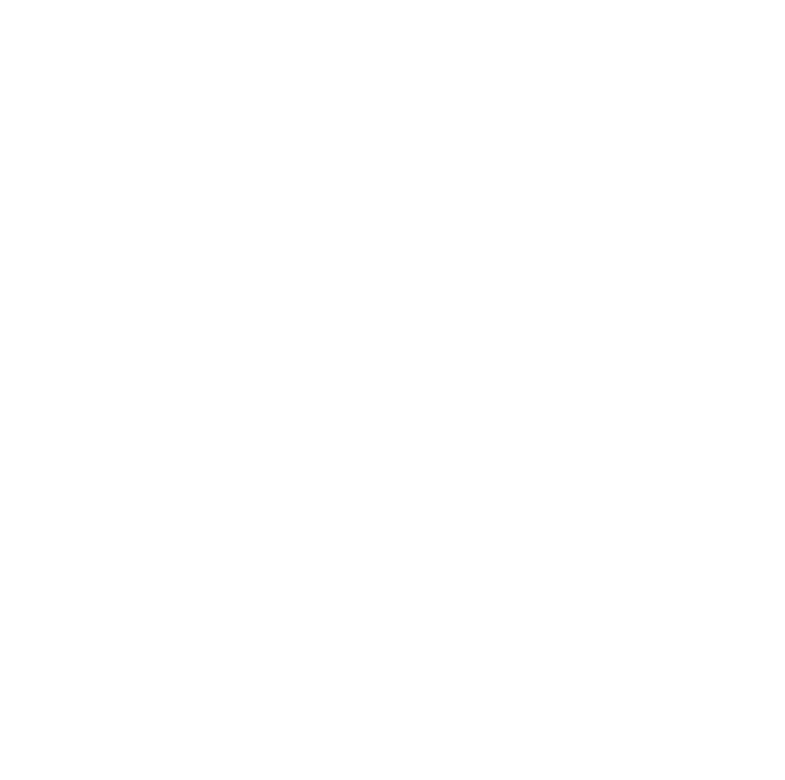}
        \\\hline\hline
    \end{tabular}
    \caption{List of concurrences corresponding to \cref{fig:concurrence}, i.e., helical state (HS), skyrmion (SK) and field-polarized (FP) state at different spin distances (nearest neighbor, next-nearest neighbor and 3rd nearest neighbor). Values are depicted in a linear gray scale, with black corresponding to the maximum in the nearest-neighbor concurrence of the HS state. We find that HS and SK states feature entangled spin pairs beyond next-nearest neighbors, whereas FP states show only minute entanglement at the system boundaries with vanishing entanglement in the bulk.}
    \label{tab:concurrence}
\end{table}
To probe whether the spins constituting a skyrmion are entangled, a genuine quantum feature of many-body systems, we compute the entanglement entropy of a suitable bipartition of the system.
The latter is defined as $S=-\tr(\rho_A\ln\rho_A)$, which can be understood as the von Neumann entropy of the reduced density matrix $\rho_A = \tr_{B}(\rho_{AB})$, obtained by splitting the set of lattice sites $N$ into two disjoint sets $A$ and $B$, and performing the partial trace over subsystem $B$.
For our purposes, it is sufficient to fix $A$ and $B$ as two patches that are symmetric about the central site (see \cref{sec:2d_1d}).

MPS are constructed to target states of small entropy by truncating the reduced density matrix to a dimension $\dim(\rho_A)\leq M$, which yields an upper bound for the entanglement entropy of an MPS state, $\tilde S=\ln M$, a quantity on the order of $1-10$ for typical simulations.
The bulk of FP states can be approximated by product states of spin-$1/2$ particles aligned to the axis of the magnetic field, and for those states, one finds $S=0$ up to small finite-size corrections.
In contrast, for systems hosting a single quantum skyrmion, we obtain values for the entanglement entropy in the range $0.2<S<0.7$, which demonstrates the presence of significant entanglement in the spin-$1/2$ quantum skyrmion and indicates that they cannot be expressed as a classical product state.
Finally, in the HS phase, we find the strongest entanglement ($0.7 < S$).
While we focus mainly on skyrmions with ferromagnetic exchange interaction ($J<0$), we find strong signatures of skyrmions and skyrmion lattices for antiferromagnetic exchange ($J>0$) as well.
However, this phase differs from the ferromagnetic skyrmions and skyrmion lattices, most apparently in the entanglement, which is significantly larger.
We postpone a detailed investigation of quantum skyrmions and skyrmion lattices for antiferromagnetic exchange couplings to future work.

\begin{figure*}[t]
    \includegraphics[width=0.32\textwidth]{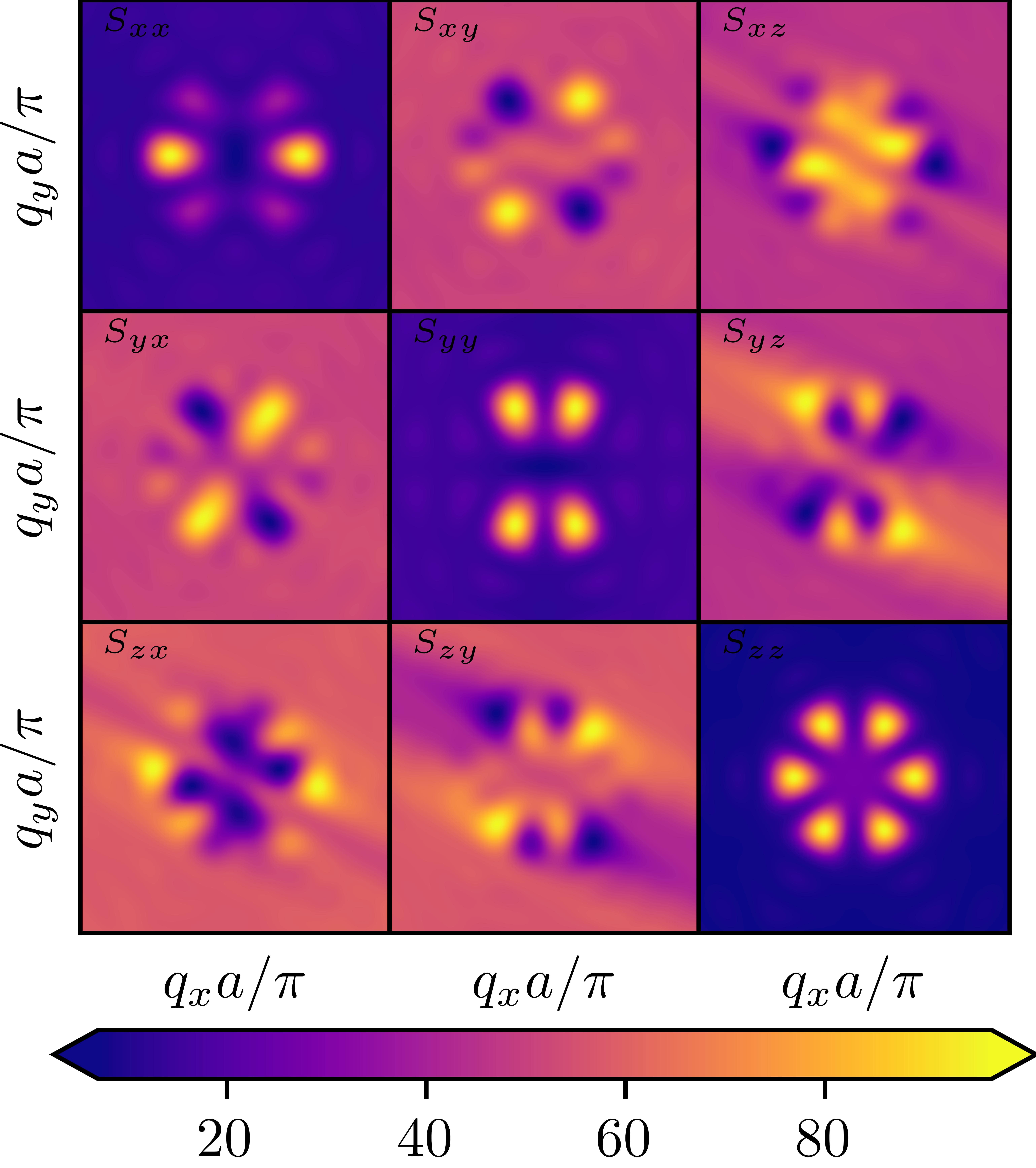}
    \includegraphics[width=0.32\textwidth]{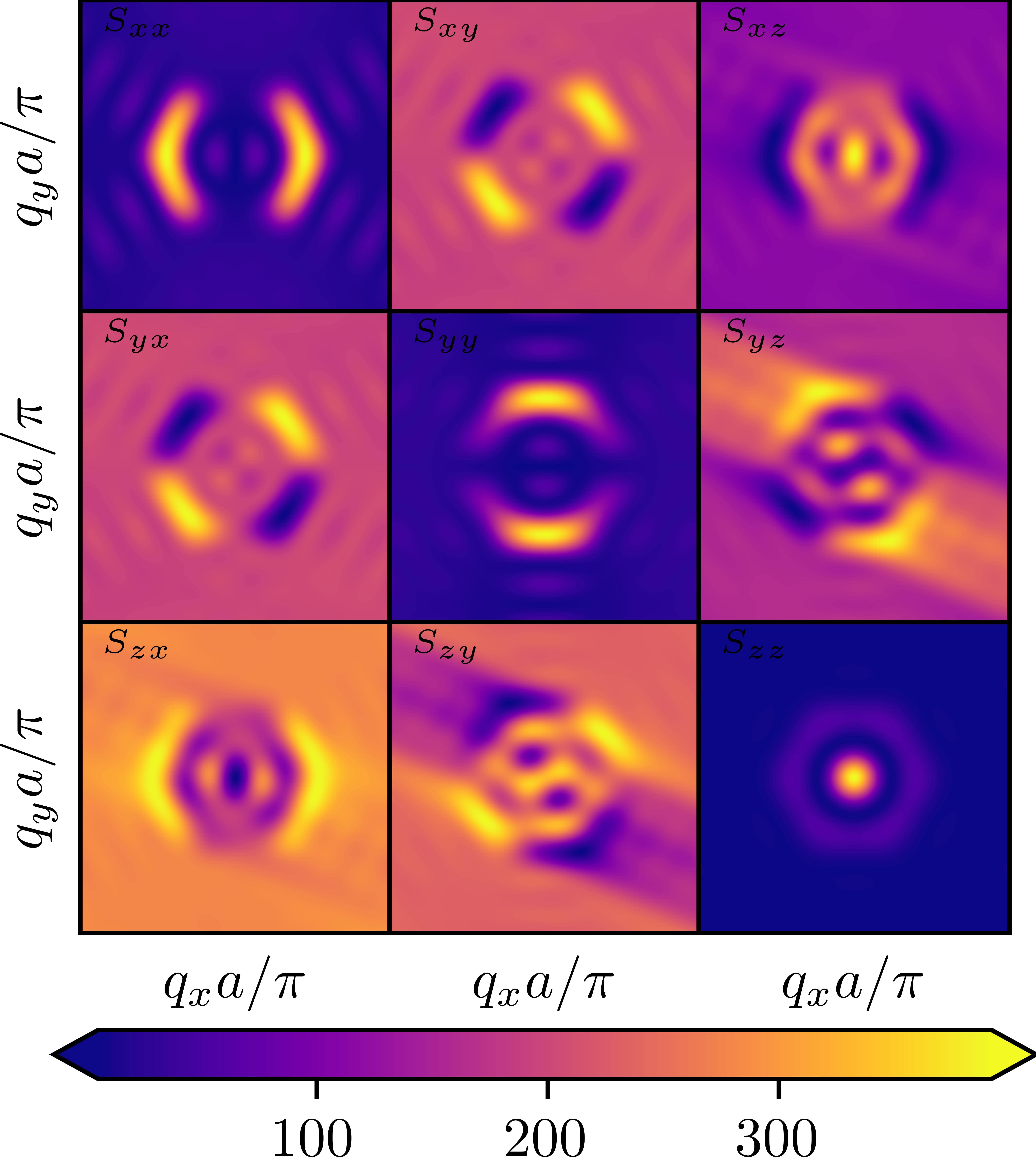}
    \includegraphics[width=0.32\textwidth]{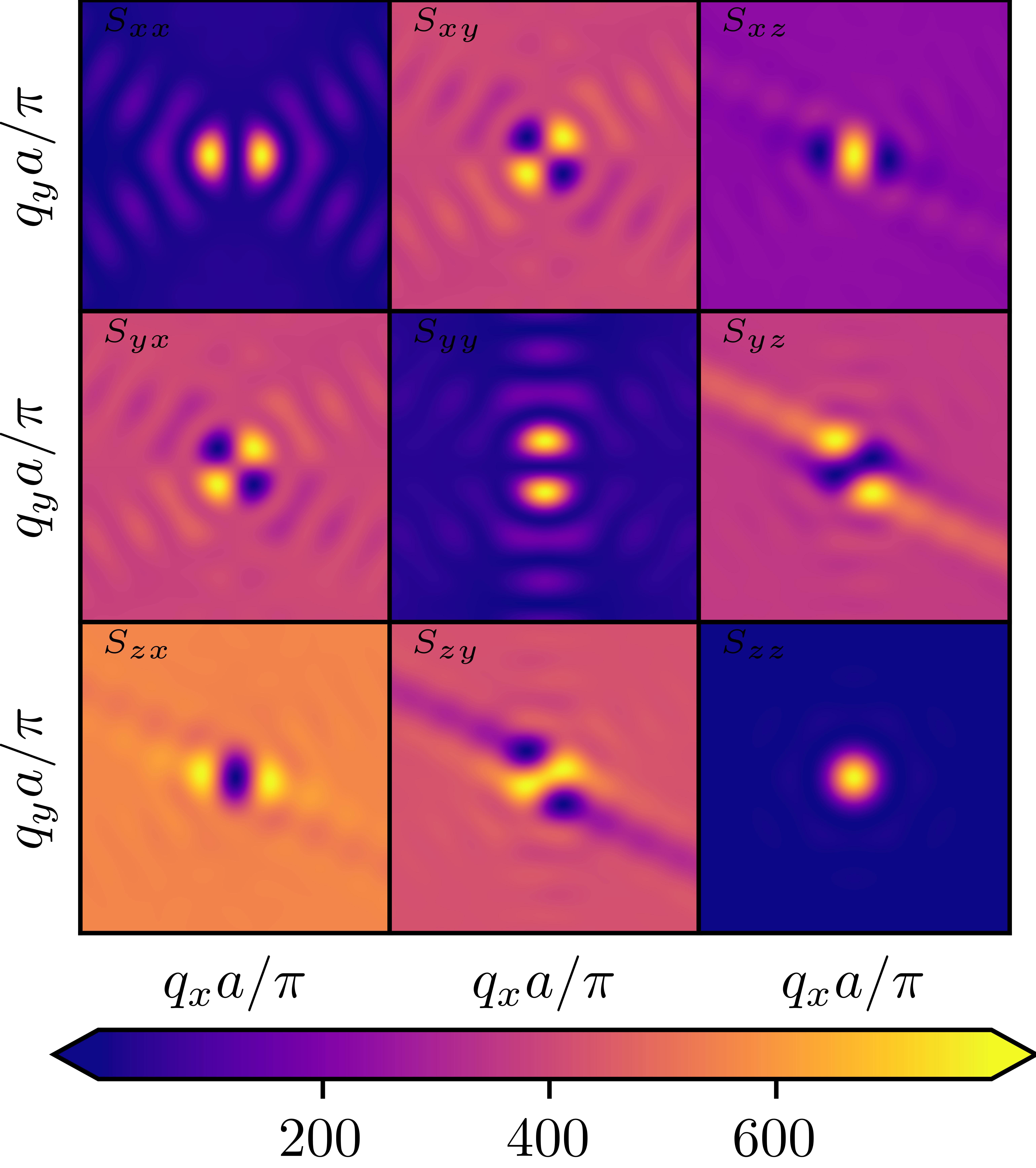}
    \caption{Components of the structure factor $\mathcal S_{\alpha\beta}(\bm q)$ (arb. units). Left:~Helical state at $B=-0.1D$. Center:~skyrmion state at $B=-0.5D$. Right:~field-polarized state at $B=-1.0D$. The geometry and local polarization is displayed in \cref{fig:three_phases_different_BC}(d)-(f).}
    \label{fig:structure_factor}
\end{figure*}
\begin{figure}[ht]
    \includegraphics[width=\columnwidth]{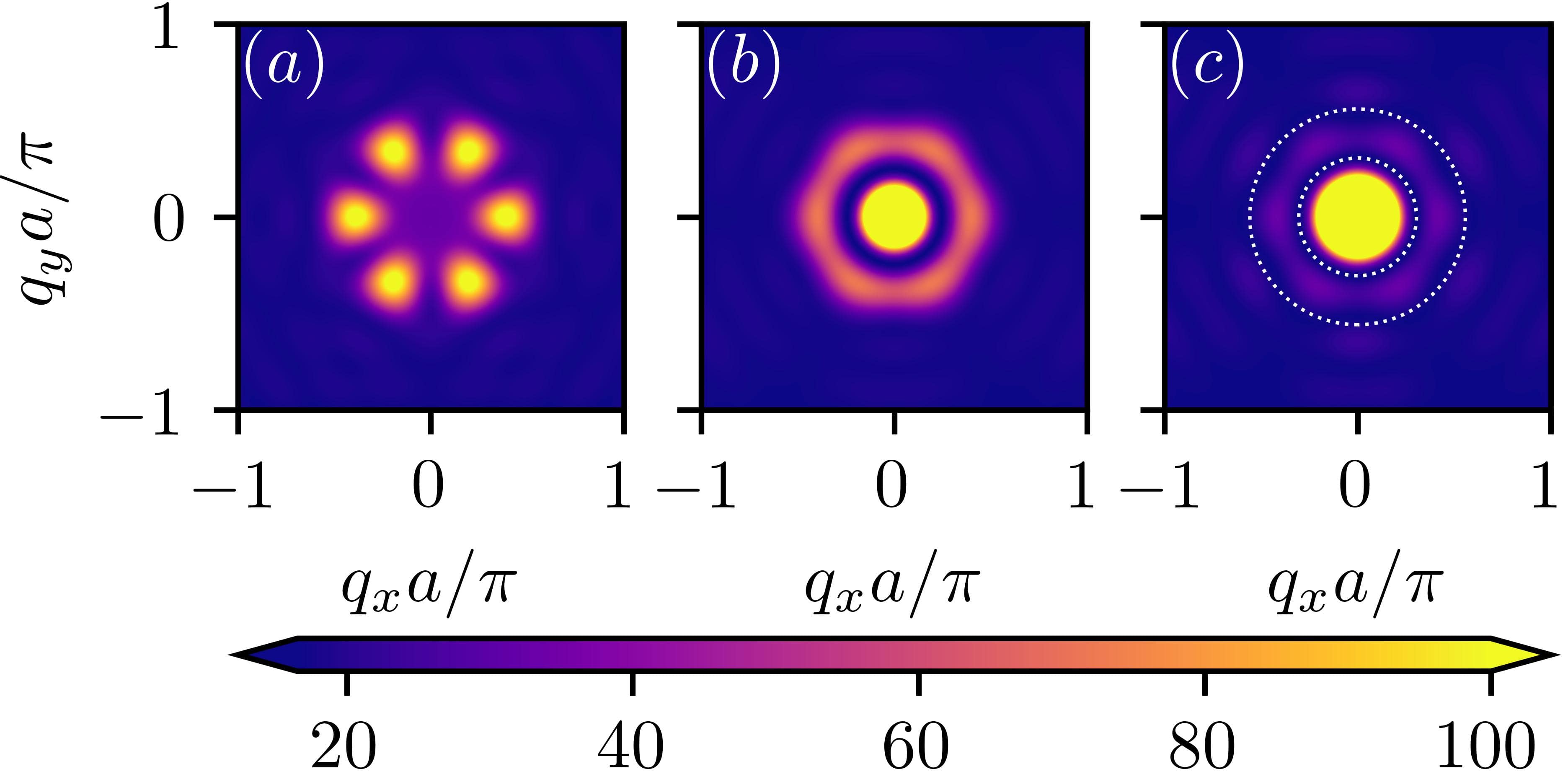}
    \caption{Elastic magnetic differential scattering cross section $d\sigma / d\Omega$ (in arbitrary units). (a)~Helical state at $B=-0.1D$, $(b)$~skyrmion state at $B=-0.5D$, and $(c)$~field-polarized state at $B=-1.0D$ for a triangular disk system of radius $R=4a$. The geometry and local polarization is displayed in \cref{fig:three_phases_different_BC}(d)-(f).}
    \label{fig:cross_section}
\end{figure}

The von Neumann entropy targets quantum correlations between a bipartition of the system but does not provide local information about the entanglement between individual spin-$1/2$ pairs.
To access the spatial distribution of entanglement, it is therefore more convenient to calculate the concurrence $C_{\bm r_1\bm r_2}$, defined for two lattice sites at positions $\bm r_1$ and $\bm r_2$.
For a generic state it can be expressed through the root of the spectrum of the non-Hermitian matrix $R_{\bm r_1\bm r_2} =  \rho_{\bm r_1\bm r_2}{\tilde \rho}_{\bm r_1\bm r_2}$, where $\rho_{\bm r_1\bm r_2} = \tr_{\bm r \notin \{\bm r_1, \bm r_2\}}(\rho)$ is the reduced density matrix of the two sites $\bm r_1$ and $\bm r_2$, and ${\tilde \rho}_{\bm r_1\bm r_2} = (\sigma_y\otimes\sigma_y)\rho_{\bm r_1\bm r_2}^*(\sigma_y\otimes\sigma_y)$ is a rotation of this reduced density matrix.
The concurrence is constructed from the square roots $\lambda_i$ (ordered in decreasing order) of the eigenvalues of $R$,
\begin{equation}
    C_{\bm r_1 \bm r_2} = \max{\{0, \lambda_1 - \lambda_2 - \lambda_3 - \lambda_4\}}.
\end{equation}
It is related to the entanglement of formation: for separable states $C$ vanishes, and it increases monotonically towards the limit $C=1$ for two maximally entangled spin-$1/2$'s~\cite{Wootters1998}.
Using the concurrence, we obtain the space-resolved entanglement distribution between spin pairs in the different phases and present its qualitative distribution by green links in \cref{fig:concurrence}.
For a more quantitative analysis, we differentiate between concurrences of different ranges up to 3rd nearest neighbor in \cref{tab:concurrence}.

For \emph{helical spin spiral (HS) states}, we find the largest concurrences up to $C_{\rm HS}=0.16$.
Therefore, we use $C_{\rm HS}$ as a measure of reference to quantify the entanglement of the remaining ordered states.
For the \emph{field-polarized (FP) states} (large magnetic field), we find almost vanishing entanglement in the bulk, indicating that the bulk spins are separable.
Small nonzero values of $C\approx 0.35C_{\rm HS}$ occur at the boundary due to the finite system size and strong DMI.
Finally, for the \emph{quantum skyrmion (SK) states} (intermediate magnetic field), we find that the spins inside the skyrmion quasiparticle are only weakly entangled, but we find concurrences $C\approx 0.8C_{\rm HS}$ at the outer rim spins of the skyrmion, signaling significant entanglement of the quantum skyrmion with the field-polarized environment.
Interestingly, both the helical state and quantum skyrmion show long-range concurrences between distant spin pairs beyond next-to-nearest neighbors.

%%%%%%%%%%%%%%%%%%%%%%%%%%%%%%%%%%%%%%%%%%%%%%%%%%%%%%%%%%%%%%%%%%%%%%%%%%%%%%%%%%%%%%%%%%%
\section{\label{sec:structure_factor}Structure factor and neutron scattering cross section}
%%%%%%%%%%%%%%%%%%%%%%%%%%%%%%%%%%%%%%%%%%%%%%%%%%%%%%%%%%%%%%%%%%%%%%%%%%%%%%%%%%%%%%%%%%%
We compute the Fourier components of the spin-spin correlation function as follows,
\begin{equation}
    \mathcal S_{\alpha\beta}(\bm q)=\sum_{\bm r \bm r'}\re^{\ri \bm q \cdot (\bm r'-\bm r)}\Braket{\hat{S}_{\vphantom\beta\alpha,\bm r\vphantom'}\hat{S}_{\vphantom\alpha\beta,\bm r'}} ,
\end{equation}
where $\alpha,\beta \in \{x,y,z\}$, $\bm q = (q_x,q_y,q_z)$ a wave vector which is later associated with the scattering vector, and the expectation values of the product of spin operators are evaluated with the ground state obtained by the MPS simulations of small hexagon flakes (panels (d)-(f) of \cref{fig:three_phases_different_BC}).
From $\mathcal S_{\alpha\beta}$, the elastic magnetic neutron scattering cross section $d\sigma / d\Omega$ at momentum transfer vector $\bm q$ is given by~\cite{loveseybook}:
\begin{equation}
    \frac{d\sigma}{d\Omega}(\bm q) \propto \sum_{\alpha\beta} (\delta_{\alpha\beta} - \hat{q}_{\alpha}\hat{q}_{\beta} )\mathcal S_{\alpha\beta}(\bm q) ,
\end{equation}
where $\hat{\bm q} = \bm{q}/q = (\hat{q}_x,\hat{q}_y,\hat{q}_z)$.
Concerning experiments, we emphasize that $d\sigma / d\Omega$ corresponds to a scattering geometry where the externally applied magnetic field $\bm B = B \hat{\bm e}_z$ is parallel to the wave vector of the incoming neutron beam, and where the detector plane is spanned by the two components $q_x$ and $q_y$ of the scattering vector.
In the limit of the small-angle approximation, one can assume $q_z \approx 0$.

In \cref{fig:cross_section} we display $d\sigma/d\Omega$ for the helical, skyrmion, and field-polarized states (see \cref{fig:structure_factor} for all components $S_{\alpha\beta}$ of the structure factor).
The geometry and local polarization of those states is displayed in \cref{fig:three_phases_different_BC}(d)-(f).
Generally speaking, long-range magnetic ordering is signaled by the presence of Bragg peaks in $d\sigma / d\Omega$ at momentum transfers $\bm q$ corresponding to the wave vectors of the ordering.
We expect additional diffuse magnetic scattering components in $d\sigma/d\Omega$ rooted in spatial variations of the spin orientation.

As expected, the cross section of the FP state in \cref{fig:cross_section}(c) is isotropic and exhibits a single broad peak centered at $\bm q = 0$ mainly caused by the component $\mathcal S_{zz}$ parallel to the external field.
The contributions in $\mathcal S_{\alpha\beta}$ ($\alpha\in\{x,y\}$, $\beta\in\{x,y,z\}$) are attributed to finite-size effects, caused by a helical winding of $\braket{\bm S_{\bm r}}$ near the boundaries due to the strong DMI interaction (see \cref{fig:three_phases_different_BC}).
The azimuthal average of the cross section in \cref{fig:cross_section}(c), defined as $(2\pi)^{-1}\int_0^{2\pi} d \varphi (d\sigma / d\Omega)$ can be well described by the form factor of a uniformly polarized thin circular disc with a radius $R$ corresponding to the cluster radius, i.e., $d\sigma / d\Omega(q) \propto [2J_1(qR)/(qR)]^2$, where $J_1(z)$ denotes the first-order Bessel function.
To highlight this point, we display the first two minima of $[J_1(qR)/(qR)]^2$ (for $R=4a$) by dotted white lines in \cref{fig:cross_section}(c) and find a very good agreement to the numerical data of the discrete system.
The HS state in \cref{fig:cross_section}(a) is characterized by a superposition of spin spirals with wave vectors $\bm q \neq 0$, resulting in six pronounced Bragg peaks in $\mathcal S_{zz}(\bm q)$.
We observe in the SK phase a superposition of the two extreme limits: in particular, we find a Bragg peak at $\bm q=0$, together with an off-diagonal Bragg ``ring'' caused by the radial polarization winding of the skyrmion (see \cref{fig:cross_section}(b)).
The radius of the skyrmion can be estimated as $r_0=q^{-1}_0\approx3a$ with $q_0\approx1/(3a)$ the momentum modulus of the $\bm q\neq0$ Bragg ring, consistent with the estimate given in \cref{fig:polarization_winding}.
Hence, the predicted quantum skyrmion profile yields a distinct signature in the measurable neutron scattering cross section and allows a determination of its size.
We want to stress that a classical spin profile compatible to the (normalized) quantum mechanical expectation values can lead to a cross section in qualitative agreement with the ones reported here.
However, quantitative deviations are expected because, for classical systems, the spin-spin correlation functions factorize since the zero temperature configuration is non-degenerate for $B\neq 0$.
Therefore, a direct measurement of the connected spin-spin correlation function $\braket{\hat\sigma_{\alpha,\bm r}\hat\sigma_{\beta,\bm r'}}-\braket{\hat\sigma_{\alpha,\bm r}}\braket{\hat\sigma_{\beta,\bm r'}}$ will differentiate classical from quantum skyrmion states.
Classical and quantum states also differ locally in the spin norm, which is not necessarily conserved in general $|\braket{\hat{\bm S}_{\bm r}}|\leq1/2$.
We find that the domain wall spin norm at the outer rim of the skyrmion (where the concurrence is large) is about $4\,\%$ lower compared to the field-polarized environment.

%%%%%%%%%%%%%%%%%%%%%%%%%%%%%%%%%%%%%%%%%%%%%%%%%%%%%%%
\section{\label{sec:skx}Quantum skyrmion lattice phase}
\begin{figure}[ht]
    \includegraphics[width=\columnwidth]{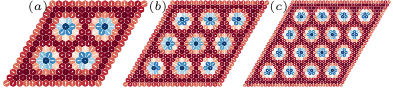}
    \caption{Local polarization of regular rhomboid triangular lattice ground states at $J=B=-0.5D$ and $K=0$. For parameters in the star-hatched region of \cref{fig:phase_diagram}(a) and larger systems, the quasiparticles are densely packed and form a skyrmion lattice.}
    \label{fig:lattices}
\end{figure}

%%%%%%%%%%%%%%%%%%%%%%%%%%%%%%%%%%%%%%%%%%%%%%%%%%%%%%%
After having discussed the properties of individual quantum skyrmions in the preceding paragraphs, we now turn to the phase diagram of the system.
Our numerical technique makes it possible to reach system sizes much larger than that of individual skyrmions, which in principle allows us to extrapolate towards a phase diagram in the thermodynamic limit.
While the HS and the FP phase remain unchanged when increasing the system size, at intermediate magnetic fields, the ground state for large lattices features a regular lattice of quantum skyrmions (SKX).
Similar to their classical analogs, the individual skyrmions form a dense packing, and for larger system sizes, we thus find quantum skyrmion chains and lattices, for which we plot examples in \cref{fig:lattices}.

\begin{figure}[ht]
    \includegraphics[width=\columnwidth]{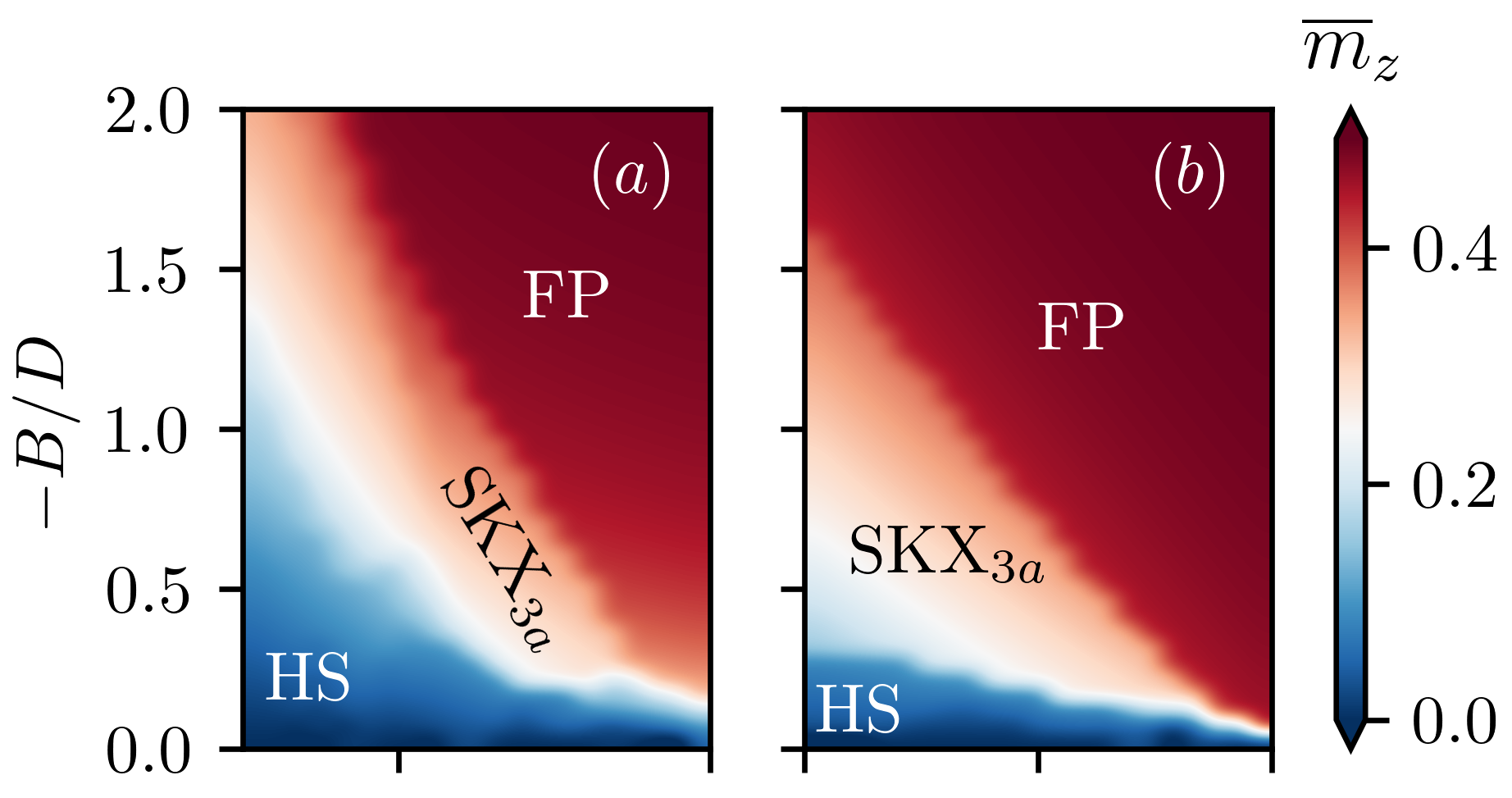}
    \includegraphics[width=\columnwidth]{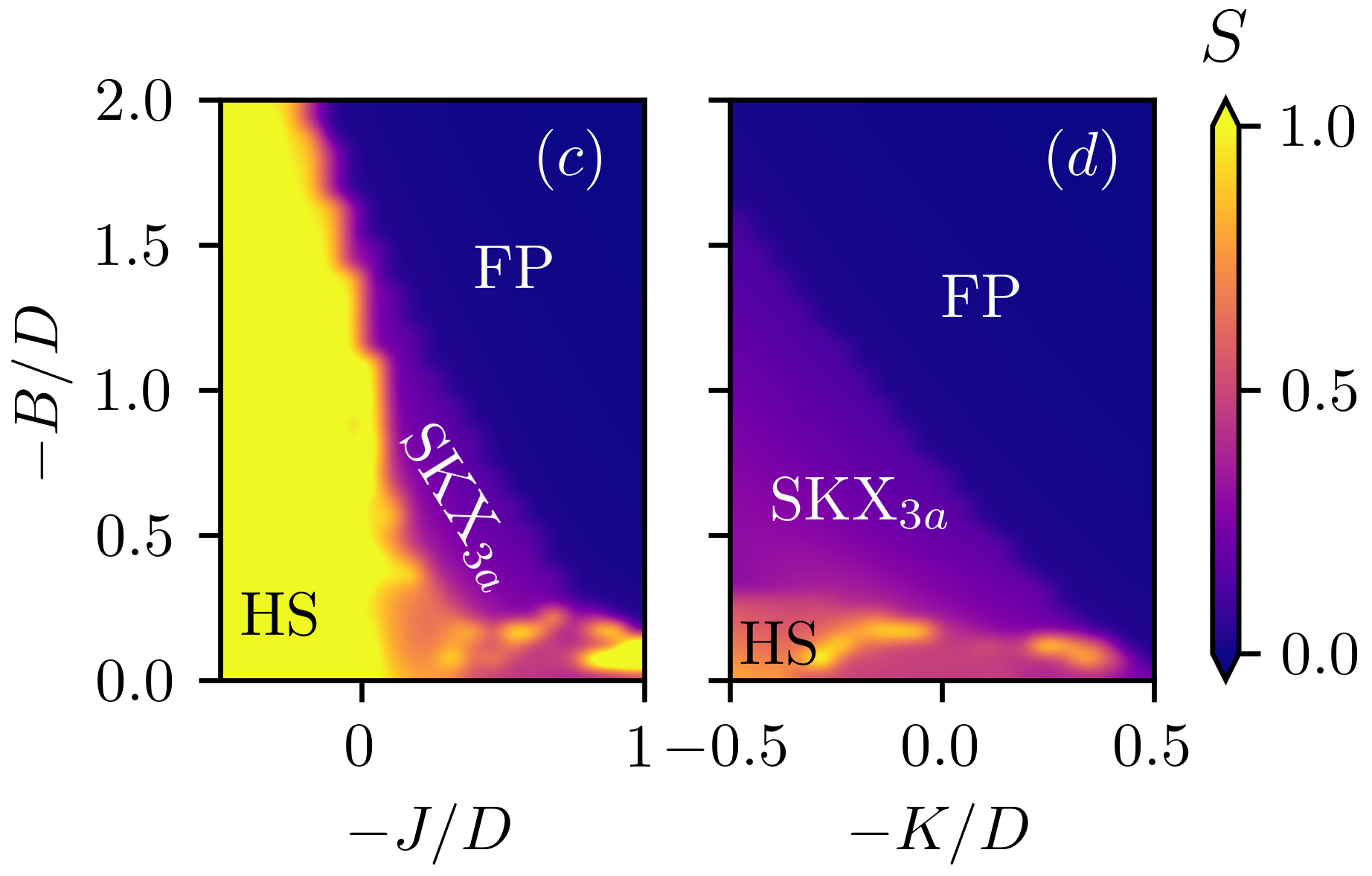}
    \caption{Average magnetization (panels $(a)$ and $(b)$) and maximum entanglement entropy (panels $(c)$ and $(d)$) of the quantum spin-$1/2$ model defined by the Hamiltonian in \cref{eq:H} and \cref{eq:HK} as a function of the external field strength $B$. The skyrmion lattice phase depicted in \cref{fig:lattices} is found within the pockets labelled ${\rm SKX}_{3a}$. The blue region hosts spin spiral states ($\rm HS$) and the red region features field-polarized ($\rm FP$) states. In $(a)$ and $(c)$, we vary $J$ with $K=0$, whereas $(b)$ and $(d)$ show the deformation of the phase boundaries by an uniaxial anisotropy $K$ with $J=-0.5D$.}
    \label{fig:phase_diagram}
\end{figure}

We elucidate the appearance and robustness of quantum skyrmion lattices as a function of the external magnetic field, the strength of DMI, and perturbations of the form \cref{eq:HK} due to uniaxial anisotropy.
For this calculation, we have concentrated on a triangular lattice with disk boundary conditions of diameter $L=9$ sites, for which we obtain a single centered skyrmion for $B=J=-0.5D$ in the unperturbed case $K=0$ (see \cref{fig:three_phases_different_BC}(e)).
We relax the fine-tuned parameter lines of \cref{fig:three_phases_different_BC} by variations of $B$, $J$, and consider nonzero uniaxial anisotropies for $J<0$ by varying $K$.

Based on our simulations, we predict the existence of three distinct quantum phases of our model: (i)~a region hosting helical spin spiral states ($\rm HS$) for weak field amplitudes, (ii)~a valley for field strengths of the order of $B\approx -0.5D$, which features a lattice formed by quantum skyrmions of radius $r\approx3a$ (${\rm SKX}_{3a}$), and (iii)~a field-polarized phase ($\rm FP$) where spins align parallel to the external field.

Helical spin spiral states are characterized by a vanishing average polarization $\overline{m}_z$, whereas field-polarized states are maximally polarized (up to finite-size effects).
As shown in \cref{fig:three_phases_different_BC}, quantum skyrmions are located in a background of field-polarized spins, and as a consequence, the state in the skyrmion lattice phase will have a finite polarization smaller than a corresponding field-polarized state.
We numerically confirm this intuitive picture and find disjoint intervals of average polarization $\overline{m}_z$ uniquely linked to each phase (see \cref{fig:polarization_plateaus}(a)), which can be summarized in the zero-temperature phase diagram presented in \cref{fig:phase_diagram}.
Note that \cref{fig:phase_diagram} is obtained by simulations of a fixed flake system size and is therefore only qualitatively correct in the thermodynamic limit.
In order to determine the exact position of quantum critical points or the nature of the quantum phase transition, a finite size extrapolation is necessary, a study which we leave for future work.

Our results about the dependence of the skyrmion phase on the uniaxial anisotropy are in qualitative agreement with corresponding classical systems, where it is known that a weak uniaxial anisotropy tends to stabilize skyrmion configurations at smaller magnetic fields~\cite{Heinze_2011,Romming_2013}.
Furthermore, we observe that the skyrmion radius increases with $-J/D$ (compare \cref{fig:phase_diagram}(e) and (h)), such that quite large spin-$1/2$ systems might be needed to resolve even individual quantum skyrmions.
Based on our results for $J=-0.5D$ we conjecture that the skyrmion lattice phase should also exist for such cases where the individual quasiparticles have a larger radius, but due to limitations dictated by the numerical complexity (which we discuss in \cref{sec:MPS}), other numerical techniques must be consulted to make quantitative predictions about the phase diagram for $-J/D\gg0.5$.
Similar values of the exchange coupling are expected in present thin film experiments~\cite{Vedmedenko2019}, thereby making our results of practical relevance.

Besides the ferromagnetic ${\rm SKX}_{3a}$ phase, which is the focus of this work, we find signatures of a quantum skyrmion lattice phase in the absence of exchange coupling and even for antiferromagnetic couplings $J\leq0$.
However, because of the significantly enhanced entanglement (see \cref{fig:phase_diagram}(c)), the MPS ansatz for this phase requires an exponential scaling of the bond dimension with the system size, and therefore simulations of large clusters are out of reach for DMRG.

%%%%%%%%%%%%%%%%%%%%%%%%%%%%%%%%%%%%
\section{\label{sec:summary}Summary}
%%%%%%%%%%%%%%%%%%%%%%%%%%%%%%%%%%%%
We have demonstrated that the ground state of the two-dimensional ferromagnetic spin-$1/2$ Heisenberg model in the presence of DMI hosts quantum skyrmions at intermediate magnetic fields $B \approx J=-D/2$.
The resulting magnetic textures are characterized by a central spin pointing opposite to the direction of the applied magnetic field and winds radially outwards towards the field-polarized environment, similar to a classical N\'eel skyrmion.
For periodic boundary conditions and in the thermodynamic limit, we expect the ground state of the skyrmion lattice phase to be degenerate, scaling with the area of the individual skyrmion quasiparticles, such that the bulk of a system with open boundary conditions corresponds to a spontaneously symmetry broken state.
The existence of quantum skyrmions yields experimental signatures in the position-dependent magnetization, the average polarization, and the structure factor, and we showed that these observables allow a distinction between a spin spiral phase at small magnetic fields, a skyrmion phase at intermediate magnetic fields, and a field-polarized phase at large magnetic fields.

While the spin texture is reminiscent of classical skyrmions, we should point out that in the present case, the skyrmion phase arises as a quantum ground state at zero temperature with open boundary conditions.
In contrast, classical skyrmions typically occur at finite temperatures and result from a minimization of the free energy.
Moreover, our examination of the resulting quantum state using the entanglement entropy and the concurrence has revealed that the quantum skyrmion state features significant entanglement shared between spin pairs of the skyrmion boundary.
We argued that the quantum and classical states can be distinguished by the norm of the polarization $|\braket{\hat{\bm S}_{\bm r}}|\leq1/2$ (conserved for classical states) and by connected correlation functions (vanishing for classical states).
We therefore conclude that a semiclassical treatment of the quantum skyrmion based on a classical magnetic texture would not necessarily capture the internal degrees of freedom of a quantum skyrmion.

Towards larger system sizes, we found that the quantum skyrmion phase is characterized by a regular lattice of skyrmions.
As the size of individual skyrmions is determined by the system parameters $B$, $J$, $D$, and $K$, a regular lattice requires commensurability between the lattice size and the skyrmion size.
While our numerical simulations cannot reach the limit of infinite system size, our results allow us to extrapolate that the ground state in the thermodynamic limit features a dense packing of quantum skyrmion textures.
Each of these quantum skyrmions has entanglement localized near its domain wall, but the entanglement between different skyrmions is small, which suggests that they can be approximated as individual quasiparticles.

We expect that our results may guide the development of an effective analytical field theory of the quantum skyrmion phase.
Based on our experience, we conclude that variational tensor networks provide a suitable numerical technique to study these systems.
This is not surprising for gapped quantum phases with short-range interactions and bounded entanglement.
Nevertheless, using MPS for a two-dimensional system is not without pitfalls, as the necessary mapping on a one-dimensional system causes non-local interactions.
We have made sure that our results have fully converged for lattice sizes corresponding to individual skyrmions.
However, we have seen that the numerical errors grow for the system sizes required for $4 \times 4$ skyrmion lattices ($29\times29$ spin-$1/2$ in total).
For such large systems, we expect our results to be only qualitatively correct.

Regarding alternative numerical schemes, we have verified that our MPS results agree quantitatively with all available results from exact diagonalization.
We have also compared our results to variational methods based on neural-network quantum states.
However, we found significant deviations between the exact result and neural-network states even for small system sizes, and the error was already on the order of 10\% for the energy eigenvalues.
This suggests that neural-network states may not provide an efficient variational ansatz for mesoscopic spin systems with DMI.
We expect that quantum Monte-Carlo simulations might be useful to go to larger system sizes.
However, the inclusion of DMI together with an external magnetic field brings about a sign problem that hinders convergence.
Other promising tensor network states for 2D spin systems with DMI are variational tree tensor network states~\cite{Shi2006} and projected entangled pair states (PEPS) \cite{Verstraete2008}.
We expect finite PEPS to outperform MPS for larger spin-$1/2$ systems hosting skyrmion lattices, especially since recently a more efficient gradient-based optimization has been developed based on automatic differentiation techniques~\cite{Liao2019,Hasik2021,liu2021gapless}.

%%%%%%%%%%%%%%%%%%%%%%%%%%%
\section*{Acknowledgements}
%%%%%%%%%%%%%%%%%%%%%%%%%%%
We gratefully acknowledge discussions with Michael Philipp Adams.
The authors acknowledge financial support from the National Research Fund of Luxembourg (FNR) under the following grants: ATTRACT A14/MS/7556175/MoMeSys, CORE C20/MS/14764976/TopRel, and CORE SANS4NCC.
This research was supported in part by the National Science Foundation under Grant No.~NSF PHY-1748958.

\appendix

%%%%%%%%%%%%%%%%%%%%%%%%%%%%%%%%%%%%%%%%%%%%%%%%%%%%%%%%%%%
\section*{\label{sec:MPS}APPENDIX A: Matrix Product States}
%%%%%%%%%%%%%%%%%%%%%%%%%%%%%%%%%%%%%%%%%%%%%%%%%%%%%%%%%%%
%
\begin{figure*}[ht]
    \includegraphics{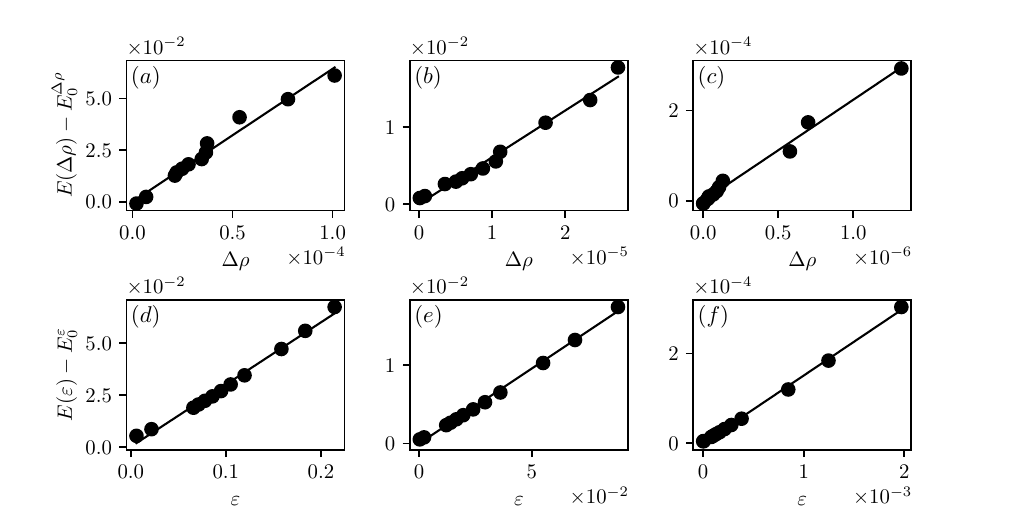}
    \caption{Energy extrapolations for the three systems plotted in \cref{fig:three_phases_different_BC}(d)-(f). Energy and error of the helical spin spiral states are displayed in panels (a) and (d), of skyrmion states in panels (b) and (e) and of field-polarized states in panels (c) and (f). The black line corresponds to a least squares fit, whose energy offset $E_0$ and error is displayed in \cref{tab:energy_extrapolation}.}
    \label{fig:energy_extrapolation}
\end{figure*}
\begin{figure*}[t]
    \includegraphics[width=\textwidth]{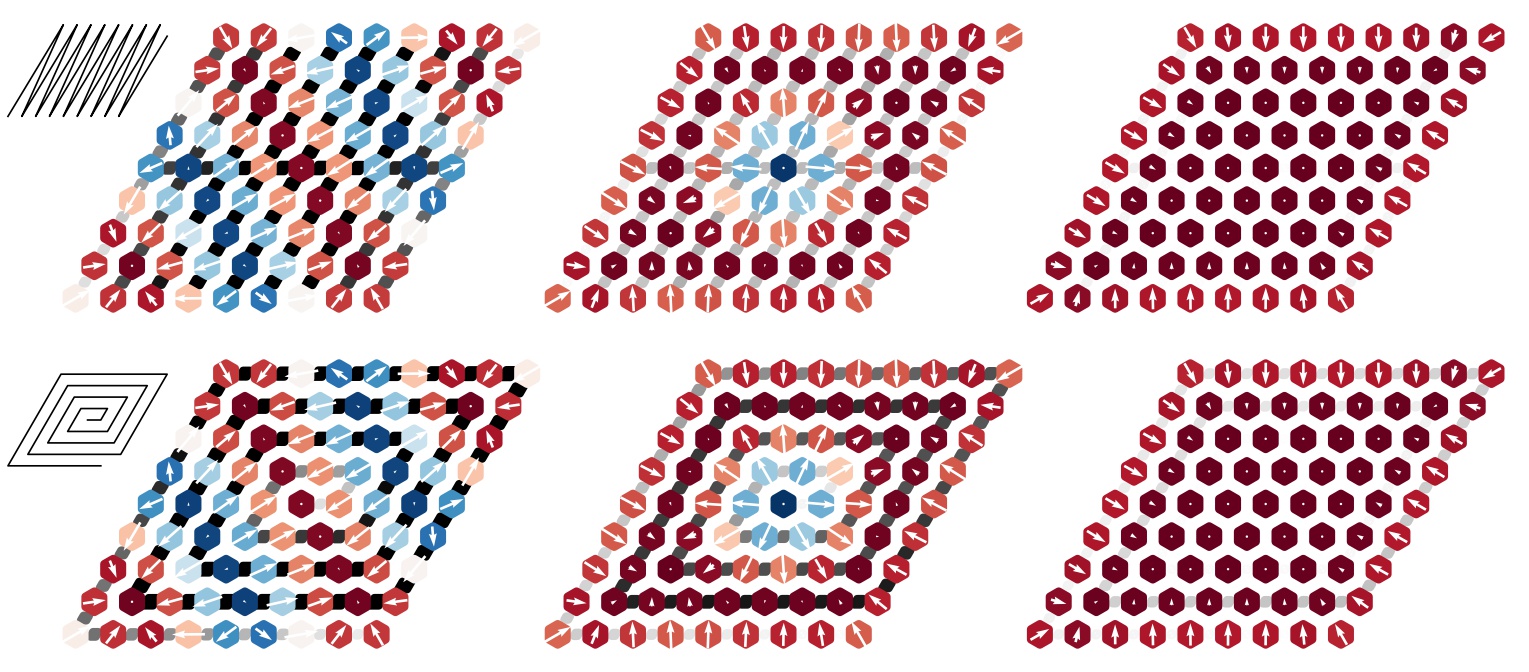}
    \caption{The three distinct ground states of \cref{eq:H} (spin spiral, skyrmion, field-polarized) obtained by two different 2D$\rightarrow$1D mapping strategies: zigzag (top row) vs. spiral (bottom row). The polarization is depicted by colors and arrows, and the von Neumann entropy is encoded by links in gray scale. Whereas the von Neumann entropy is rather small and homogeneously distributed for the zigzag order, it appears not only larger but also inhomogeneous for the spiral ordered 1D chain.}
    \label{fig:zigzag_vs_spiral}
\end{figure*}
\begin{figure}[htbp]
    \includegraphics[width=.475\columnwidth]{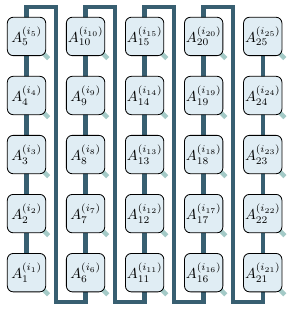}
    \hfill
    \includegraphics[width=.475\columnwidth]{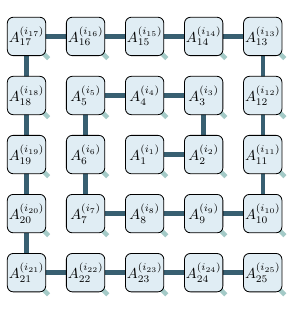}
    \caption{Two distinct mappings of the underlying 2D spin lattices in terms of a 1D MPS representation. Dark blue lines represent contractions over the auxiliary (bond) dimension of the tensors, while diagonal lines indicate the physical dimension $i_n$.}
    \label{fig:mps_order}
\end{figure}
Tensor networks provide an important numerical toolbox in computational physics and have been applied successfully to countless interacting and strongly correlated systems~\cite{cirac2020,Bauls2020,Baiardi2020,Paeckel2019}.
One of the most established algorithms, called DMRG~\cite{White1992,Hubig2015,NunezFernandez2020}, is understood as a sequential variational optimization of adjacent MPS tensors until convergence is reached.
Our DMRG simulations are mainly based on the Julia package \href{https://github.com/ITensor/ITensors.jl}{ITensors}~\cite{itensors}, and we made available a condensed version, reproducing \cref{fig:three_phases_different_BC}(e), on GitHub \cite{Haller2021Zenodo}.
Additionally, we cross-checked ITensor with \href{https://tenpy.readthedocs.io/en/latest/}{TeNPy} simulations \cite{Hauschild2018}, and upload the condensed version for the TeNPy framework alongside the julia implementation \cite{Haller2021Zenodo}.

A generic state consisting of $N$ spin-$1/2$ sites reads
\begin{equation}
    \ket\psi = \sum_{i_1,i_2,\dots,i_N = \uparrow,\downarrow} \left(\prod_{n=1}^N A^{(i_n)}_n\right) \ket{i_1,i_2,\dots,i_N},
\end{equation}
where $\{\ket{i_k}\}$ forms a canonical basis of the Hilbert space at site $k$ out of $N$ sites in total.
For a finite system with Dirichlet boundary conditions, the objects $A^{(i_n)}_n$ are matrices, except for the two boundary vectors $A^{(i_1)}_1$ and $A^{(i_N)}_N$, such that the result of the product is a scalar.
The dimension of the matrices $M=\max_n \dim(A^{(i_n)}_n)$ is called the bond dimension.
If $M$ is fixed to an arbitrary integer, the MPS representation of quantum states can be used as a variational ansatz to approximate the minimum energy eigenstate.
The quality of this approximation is controlled by $M$.
This is particularly transparent in the so-called Schmidt decomposition $\ket\psi = \sum_{i=1}^M s_i \ket{\psi_{A,i}}\ket{\psi_{B,i}}$ in which $A$ and $B$ denotes an arbitrary bipartition of the system and $\ket{\psi_{\alpha,i}}\in\mathcal H_\alpha$ forms a complete basis in the Hilbert space of the part $\alpha\in\{A,B\}$.
The Schmidt values $s_i$ are the roots of the eigenvalues of the reduced density matrix, and therefore related to the von Neumann entanglement entropy $S=-\sum_{i=1}^M |s_i|^2\ln(|s_i|^2)$.
Consider a truncation $\tilde M<M$, then states with a small weight in the reduced density matrix are neglected, and the overlap between the original state with bond dimension $M$ is reduced.

Similar to quantum states having an MPS representation, quantum operators have a matrix product operator (MPO) representation.
Let $M_H$ be the bond dimension of the Hamiltonian MPO, then the standard DMRG algorithm bears a leading numerical complexity of $\mathcal O(M^3M_H)$ (assuming that $M_H^2<M^2$).

For generic many-body states rewritten as MPS, $M$ is an extensive quantity in the number of sites and diverges in the thermodynamic limit.
If the target state of a one-dimensional system obeys an area law of the quantum entanglement, the von Neumann entropy is guaranteed to be a finite constant~\cite{Hastings2007}.
Consequentially, $M$ remains finite in the thermodynamic limit, and MPS becomes exact, which explains the success of DMRG applied to one-dimensional quantum systems.
Despite its limitations in two dimensions, DMRG is frequently applied to ladder systems and can even yield reliable results for strongly correlated lattices, especially in the case of quantum spin-$1/2$ Heisenberg models.
We expect MPS to reliably capture the physics of the quantum skyrmion lattice phase because the external field polarizes the environment, and the resulting states carry no entanglement in the paramagnetic regions, {\it but localized entanglement around the domain wall of the skyrmion}.

We typically start with random MPS initial states of bond dimensions up to $M\leq1024$, followed by sequential variational optimizations (``sweeping'') of two adjacent tensors (two-site DMRG).
The two-site DMRG allows us to estimate the truncation error $\Delta\rho = \sum_{i=M+1}^\infty s_i^2$, which we use in \cref{fig:energy_extrapolation} to extrapolate towards results without numerical errors.
To ensure that we display converged results only, we carefully monitor local spin expectation values and stop the simulation if changes in the observables become smaller than $\delta = 10^{-10}$.
Since we use DMRG in two spatial dimensions, convergence to a spin spiral state may require many sweeps, on the order of $100-1000$.
The quality of the approximate ground state with energy $E(M)=\braket{\psi(M)|\hat H|\psi(M)}$ can be estimated by the energy variance
\begin{equation}
    \varepsilon(M) = \Braket{\psi(M)|\left(\hat H - E(M)\right)^2|\psi(M)} .
    \label{eq:error}
\end{equation}
By construction, $\varepsilon = 0$ for exact eigenstates of the Hamiltonian.
Since MPS approximates the wave function with a finite bond dimension $M$, we have $\varepsilon(M)>0$ and $\lim_{M\rightarrow\infty}\varepsilon(M)=0$ in general.
Similarly, $\lim_{M\rightarrow\infty}E(M)=E_0$ converges to the true eigenstate energy.
To estimate the numerical error of our approximate wave functions, we perform linear extrapolations of the energy scaled against the truncation error~\cite{White2007} and the energy variance~\cite{Saadatmand2016}.
The outcomes of this extrapolation are presented in \cref{fig:energy_extrapolation} and \cref{tab:energy_extrapolation}.
\begin{table}[t]
    \centering
    \begin{tabular}{|l|l|l|l|}
        \hline
        $B/D$ & $-0.1$ & $-0.5$ & $-1.0$ \\
        \hline\hline
        $E^{\Delta\rho}_0$ & $-34.85093(9)$ & $-41.024253(4)$ & $-53.428962927(1)$ \\
        $E^{\varepsilon}_0$ & $-34.85707(4)$ & $-41.024009(1)$ & $-53.4289732208(1)$ \\
        \hline\hline
    \end{tabular}
    \caption{Least squares energy fit for the data of \cref{fig:energy_extrapolation}.}
    \label{tab:energy_extrapolation}
\end{table}
We want to stress that the MPS approximations corresponding to skyrmion and field-polarized states easily reach convergence within a few dozen sweeps and follow the expected linear trend in the approximation errors $\Delta\rho$ and $\varepsilon$ \cite{Haller2021Zenodo}.

Changes in the local spin expectation values beyond $M=128$ are invisible to the naked eye when displayed on the scales used in the main text such that a detailed error extrapolation is not needed.
A word of caution is due in the case of helical spin spiral states:
as we already explained in the main text, these states are difficult to simulate using MPS due to the large degeneracy of the ground state manifold.
This leads to some issues in reaching convergence (up to $1000$ sweeps are needed) which for too small bond dimensions may even cause DMRG to get stuck in local energy minima corresponding to excited eigenstates.
In the helical phase and for large lattices, MPS is thus not always reliably converging to approximations of the global ground state but converges under some circumstances to low-lying excited states with less entanglement -- with outcomes roughly comparable with those presented in Ref.~\cite{Jiang2012natcom}.

%%%%%%%%%%%%%%%%%%%%%%%%%%%%%%%%%%%%%%%%%%%%%%%%%%%%%%%%%%%%%
\section*{\label{sec:2d_1d}APPENDIX B: Mapping from 2D to 1D}
%%%%%%%%%%%%%%%%%%%%%%%%%%%%%%%%%%%%%%%%%%%%%%%%%%%%%%%%%%%%%
Before we can apply DMRG to the system at hand, the 2D lattice must be mapped to a 1D chain.
The map from a 2D lattice to a 1D chain can be performed by a sequential numbering of the lattice nodes with major ordering along an arbitrary axis (zigzag order).
We choose the major axis to be ${\bm a}_2$.
This can be achieved by $f(\bm r(n_1,n_2)) = n_2 + \sum_{n<n_1}l(n)$, where $l(n)$ is an auxiliary function that encodes the lattice open boundary conditions and $\bm r(n_1, n_2) = \sum_i n_i{\bm a_i}$.

As a result, the lattice Hamiltonian $\hat H = \sum_{\langle\bm r, \bm r'\rangle}\hat H_{\bm r,\bm r'} + \sum_{\bm r}\hat H_{\bm r}$ is mapped to a chain Hamiltonian $\hat H = \sum_{\langle\bm r, \bm r'\rangle}\hat H_{f(\bm r),f({\bm r}')} + \sum_{\bm r}\hat H_{f(\bm r)}$.
To simplify the remaining discussion, we now assume square or rhomboid boundary conditions (see \cref{fig:mps_order}), and $1<n_i<L_i$, such that $l(n)=L_2$.
On-site contributions remain local, nearest-neighbor interactions along the major axis remain short ranged, but the interactions along the ${\bm a}_1$ axis now have an extended range $|f(\bm r) - f({\bm r}\pm{\bm a}_1)| = L_2$.
This results in a growth of the dimension of the Hamiltonian matrix product operator $M_H\propto L_2$.
To obtain reasonable computation times for large skyrmion lattice systems, one must therefore restrict the bond dimension $M$ to significantly smaller values.
For the largest quantum skyrmion lattice system we present in \cref{fig:lattices}, we plot the converged results of $M=128$.
Note that the choice of our mapping preserves the locality of the interaction in one direction.
In an attempt to remove this bias, we checked the resulting MPS quality for a different mapping, starting at the central spin-$1/2$ site and ordered radially outward (spiral ordering).
Using the spiral ordering, we can confirm using the von Neumann entropy that the outer rim of the skyrmion is strongly entangled with its environment, a conclusion we had also reached based on the concurrence.

Compared to the other proposition, the spiral mapping results in higher variational energy, likely the result of the inhomogeneous entanglement distribution (see \cref{fig:zigzag_vs_spiral}).
Since entanglement can be created by non-local transformations, it is known that certain mappings from 2D to 1D are beneficial compared to others, which can be utilized to obtain a substantial improvement of the overall simulation quality~\cite{Cataldi2021}.
For the results presented in the main text, we consistently use the zigzag order.
The phase diagram presented in \cref{fig:phase_diagram} is entirely unaffected by this choice.

%%%%%%%%%%%%%%%%%%%%%%%%%%%%%%%%%%%%%%%%%%%%%%%%%%%%%%%%%%%%%%%%%%%%%%%%%%%%
\section*{\label{sec:classical_vs_quantum}APPENDIX C: Classical vs. Quantum}
\begin{figure}[ht]
    \centering
    \includegraphics{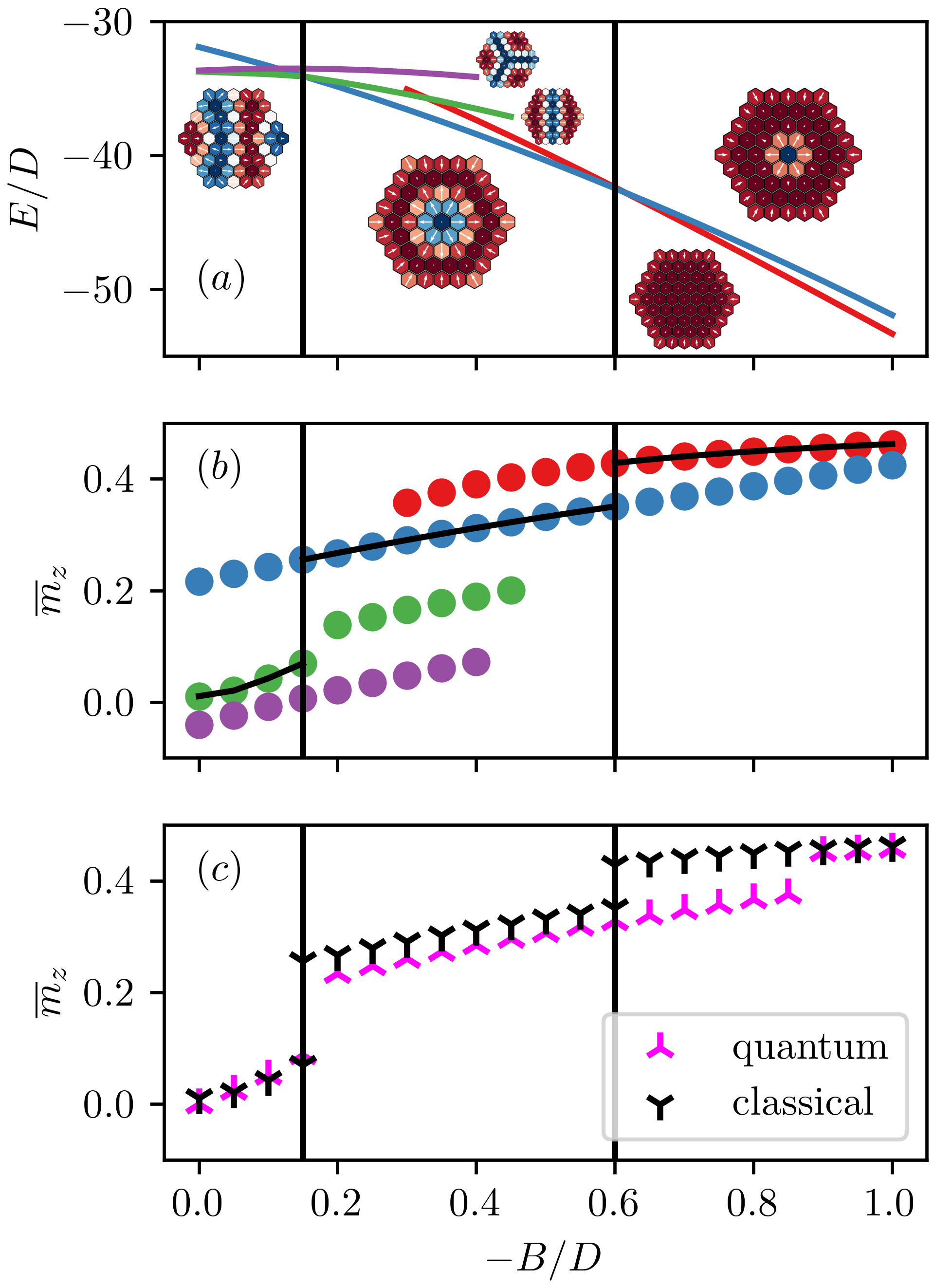}
    \caption{(a) Classical energy and (b) magnetization obtained by variational optimization (see text). Different colors correspond to different configurations, which are presented close to their energy lines in panel (a). In (c), we compare the average magnetization as a function of $B/D$.}
    \label{fig:classical_energy_magnetization}
\end{figure}

%%%%%%%%%%%%%%%%%%%%%%%%%%%%%%%%%%%%%%%%%%%%%%%%%%%%%%%%%%%%%%%%%%%%%%%%%%%%
A basic understanding of the classical low-energy configurations can be achieved by performing a variational minimization of the energy functional.
In particular, we want to solve for the minimum energy spin configuration which satisfies
\begin{align}
    E_{\rm min}=\min_{\{\bm S_i\ \forall i=1,...,N\}}E(\bm S_1, \bm S_2, \dots, \bm S_N)
\end{align}
in which $E$ is the classical energy functional
\begin{align}
    E &= \frac12\sum_{\braket{\bm r, \bm r'}}\left[J{\bm S}_{\bm r} \cdot {\bm S}_{\bm r'} + {\bm D}_{{\bm r}'-{\bm r}}\cdot \left( {\bm S}_{\bm r}\times{\bm S}_{\bm r'}\right)\right] + \sum_{\bm r}{\bm B} \cdot {\bm S}_{\bm r}
\end{align}
sharing the same notational conventions with the quantum Hamiltonian, except the use of classical spins.
We implemented a standard variational optimization with the \href{https://juliahub.com/ui/Packages/Optim/R5uoh/1.7.2}{Optim} julia package \cite{Mogensen2018}, which is included in our repository \cite{Haller2021Zenodo}.
All variational techniques are prone to being trapped in local minima, which sensibly depends on the initial state.
To be sure that for small $61$-spin flakes we obtain samples of a global minimum energy configuration, we performed the variational optimization with $1000$ different initial states where the azimuth and polar angles are sampled with a uniform distribution.
The low energy results of the classical setup are obtained analogously to the quantum case: we analyze the lowest energy configurations as a response to a changing external Zeeman field, for which we fix the norm of the classical spins to $1/2$.
We present a condensed version of the classical low energy results in \cref{fig:classical_energy_magnetization}.

We find that the spectral energy and magnetization lines of the different configurations as a function of $B/|D|$ are continuous over a wide range of parameters in the phase diagram.
In particular, the spectral lines show crossings, which we identify with a phase transition: the first excited states become ground states and vice versa, causing the sudden jumps in the magnetization.
This simple analysis suggests first-oder Zeeman field induced phase transitions in this model, which are conjectured to be present in the quantum case as well.

If we compare the ranges of the skyrmion phase between quantum and classical, we note that the quantum skyrmions are ground states in the regions of the classical field polarized states, which is in agreement with the results presented in \cite{RoldanMolina2015}, namely that quantum fluctuations stabilize skyrmion textures.

\bibliography{biblio,TensorNetworks_2DHeisenberg}

\end{document}